\documentclass[%
reprint,
 amsmath,amssymb,
 aps,
 prx,
superscriptaddress,
]{revtex4-1}

\usepackage{indentfirst}  
\usepackage{bm}    
\usepackage{graphicx}  
\usepackage{latexsym}
\usepackage{graphicx}
\usepackage{psfrag}
\usepackage{epsfig}
\usepackage{amsmath}
\usepackage{tocloft}
\usepackage{array}
\usepackage{setspace}
\usepackage{amssymb}
\usepackage{float}
\usepackage{multirow}
\usepackage{slashbox}
\usepackage{color}
\usepackage{hyperref}
\usepackage[config=altsf]{subfig}
\usepackage{booktabs}
\usepackage{fullpage}

\begin{document}

\preprint{APS}

\title{Ultra-low-frequency radio waves as signals and special electromagnetic counterparts of gravitational waves (from binary mergers) having tensorial and possible nontensorial polarizations}


\author{Hao Wen}
\email[]{wenhao@cqu.edu.cn}
\affiliation{Physics Department, Chongqing University, Chongqing 401331, China.}

%
%
%
%

\date{\today}

\begin{abstract}
\indent 
{\color{black} 	
Gravitational waves (GWs, from binary merger) interacting with super-strong magnetic fields of the neutron star (in the same binary system), would lead to perturbed electromagnetic waves [EMWs, in the same frequencies of these GWs,  partially in the ultra-low-frequency (ULF) band for the EMWs]. Such perturbed ULF-EMWs are not only the signals, but also a new type of special EM counterparts of the GWs. 
Here, generation of the perturbed ULF-EMWs is investigated for the first time, and the strengths of their magnetic components are estimated to be around $10^{-12}$Tesla to $10^{-17}$Tesla (in fISCO) at the Earth for various cases [not including the influence of interstellar medium (ISM)].
The components with higher frequencies of the ULF-EMWs (e.g., especially produced by the GWs of the post-merger stage) above 1.8kHz (typical plasma frequency around solar system in the Milky way), could propagate through the  ISM from the source until the Earth, and the perturbed ULF-EMWs will be reprocessed before they arrived at the Earth due to the ISM.
Also, the waveforms of the perturbed ULF-EMWs will be modified into shapes different but related to the waveforms of the GWs, by the amplification process during the binary mergers which could amplify the magnetic fields into $10^{12}$Tesla or even higher.
Specific connection relationships between the polarizations of the perturbed ULF-EMWs and the polarizations (tensorial and possible nontensorial) of the GWs of binary mergers, are also addressed here.
Characteristic properties of the perturbed ULF-EMWs (which would bring us some different new information of fundamental properties of the gravity and Universe) will be very helpful for \mbox{extracting} the signals from background noise for possible observations in the future.\\}

\begin{description}
\item[PACS numbers]
04.30.-w, 04.50.Kd, 04.50.+h, 04.50.-h
\item[Keywords]
\end{description}
\end{abstract}

                              
\maketitle

\section{Introduction}
\label{Introduction}
\begin{figure*}[htp]
	\centerline{\includegraphics[width=7.1 in]{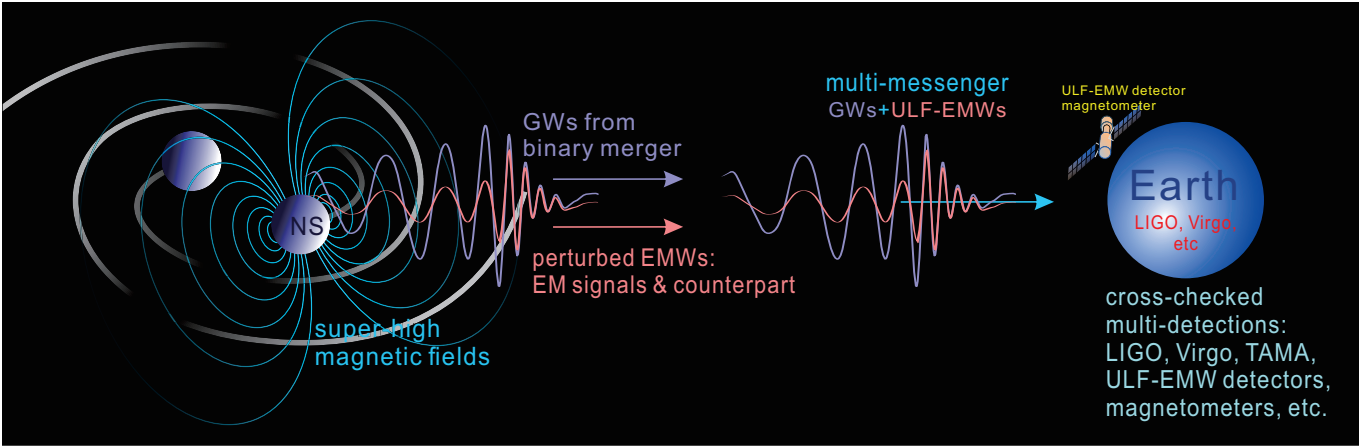}}
	\begin{spacing}{1.2}
		\caption{\footnotesize{\textbf{A general frame:}
				if one star in the binary merger is a neutron star, or even a magnetar, the binary system (i.e., neutron star-neutron star binary, or black hole-neutron star binary) could have ultra-strong magnetic fields up to $\sim10^{11}$Tesla. Moreover, during the amplification process, the magnetic fields would be greatly amplified and could easily reach $10^{12}$Tesla or much higher. The GWs from such binary merger could interact with such extremely high magnetic fields, and lead to the perturbed EMWs [also in the GW frequency band, and usually defined as ultra-low-frequency (ULF) band in the context of EMW researches], which propagate to far field area until the Earth with the GWs. Such perturbed ULF-EMWs could be a new type of signals and special EM counterparts of GWs from binary mergers, and they would have characteristic waveforms and particular polarizations which may reflect different new information about some crucial issues including the tensorial and possible nontensorial polarizations of GWs from binary mergers, magnetic fields in the mergers, etc.}}
		\label{onelayerbi}
	\end{spacing}
\end{figure*}

\indent The LIGO scientific collaboration and the Virgo collaboration have so far reported multiple gravitational wave (GW) events (e.g. GW150914, GW151012, GW151226, GW170104, GW170608, GW170729, GW170809, GW170814, GW170817, GW170818, GW170823) \cite{PhysRevLett.116.061102,secondLIGOGW,PhysRevLett.118.221101,GW170608,GW170814,bns,LIGOnew1,PRX041015} from binary black hole mergers \cite{PhysRevLett.116.061102,secondLIGOGW,PhysRevLett.118.221101,GW170608,GW170814,LIGOnew1,PRX041015} (with frequencies around 30Hz to 450Hz and dimensionless amplitudes $\sim10^{-21}$ to $\sim10^{-22}$ near the Earth) or from  binary neutron star merger \cite{bns} [GW170817, comes with the first electromagnetic (EM) counterpart of the GWs]. 
These great discoveries inaugurated a new era of GW astronomy.   
 Meanwhile,  the EM counterparts which may occur in association with corresponding observable GW events, have also been massively studied based on various emission mechanisms\cite{2041-8205-826-1-L6,AA2012,AA2016,Greiner,Smartt,0067-0049-225-1-8,2041-8205-826-2-L29,liuyuxiao,Nissanke,Kocsis,PhysRevD.81.084007,Lazzati,Lamb,0004-637X-835-1-103} due to  that they can bring us crucial information with rich scientific values.\\
\indent Further, in order to obtain more extensive and in-depth astronomical information, on the path forward for multi-messenger detections, 
  it will be very expected to expand the observations in broader frequency bands (low, intermediate, high, and very high-frequency bands), through a wider variety of principles  and methods with different effects,  aiming on more types of sources, to explore richer information of properties of the gravity and Universe. E.g., some interesting questions would arise:  how can we find the possible nontensorial polarizations of GWs predicted by gravity theories beyond GR including those with extra-dimensions? 
  Can we acquire any new EM counterparts generated by other mechanisms with different characteristics?
 
\indent In this article we address a novel topic relevant to above questions. 
A possible new EM signal of GWs from the binary mergers is proposed here, based on the effect of the EM response to GWs, which had been long studied\cite{Braginsky.Grishchuk.1973,Boccaletti_NuovoCim70_1970,prd2915,Chen1995PRL,Chen1994,LONG1994382,FYLi_PRD67_2003,FYLi_EPJC_2008,FYLi_PRD80_2009,Li.Fang-Yu.120402,Li2011,WEN2019114796,PRD104025,WenEPJC2014,LiNPB2016,wenCPC2017,PhysRevD.98.064028,Li.Wen.arXiv1712.00766,PhysRevD.94.024048}, but were usually in very high frequency bands such as over GHz ($10^9$Hz)\cite{FYLi_PRD67_2003,FYLi_EPJC_2008,FYLi_PRD80_2009,Li.Fang-Yu.120402,Li2011,WEN2019114796,PRD104025,WenEPJC2014,LiNPB2016,wenCPC2017,PhysRevD.98.064028,Li.Wen.arXiv1712.00766}.
	Whereas, we now apply such mechanism targeting on the GWs in the intermediate band (around $\sim 10^1$ to $\sim10^3$Hz) from the binary mergers. \\
\indent Specifically, as demonstrated in Fig. \ref{Introduction}, considering at least one star in the binary is a neutron star (or magnetar, which usually has ultra-strong surface magnetic fields $\sim10^{11}$Tesla), according to the electrodynamics in curved spacetime, during the binary merger, the produced GWs could interact with such ultra-strong  magnetic fields of the same source, and then lead to significant perturbed EMWs in the same frequency band [partially within the  ultra-low-frequency (ULF) band ($\sim10^2$ to $\sim10^3$Hz) in the context of researches for EMWs].
These perturbed ULF-EMWs (as a new type of signals and EM counterparts of GWs from binary mergers) and the GWs start at the same time and then propagate outward (however, their paths can be different depending on the large scale structure of the spacetime).
Such multi-messenger signals of GWs+ULF-EMWs could be observed via different corresponding methods.
Particularly, we should notice that for some cases, especially for the merger/post-merger stage of binary, the perturbed ULF-EMWs have frequencies above typical plasma frequency (e.g., 1.8KHz) of the interstellar medium  (ISM), and thus they can propagate via the ISM until the Earth (see details in Sect.\ref{estimation}). However, we here do not include effects of some other influence of the ISM, such as the dispersion, scattering, etc. These specific experimental issues for practical astrophysical observations are much more complicated and will not be the points focused in this article of theoretical proposal and estimations. If consider other effects of the ISM, the perturbed ULF-EMWs will be reprocessed during the propagation, and relevant topics will be addressed in separated works elsewhere.\\ 

\indent For the first step, instead of massive numerical computing, we apply a typical model\cite{PhysRevD.64.083008} of surface magnetic fields of neutron stars for calculation, to try to obtain a primary estimation of the order of magnitude of the signal strengths. 
Based on the electrodynamics equations in curved spacetime and previous works\cite{Li.Wen.arXiv1712.00766,PRD104025,WenEPJC2014,FYLi_PRD80_2009,FYLi_EPJC_2008,Boccaletti_NuovoCim70_1970},  the perturbed ULF-EMWs are estimated for various cases including that having tensorial and possible nontensorial GWs, and the strengths of their magnetic components would be generally around {\color{black}$10^{-12}$Tesla to $10^{-17}$}Tesla at the Earth (not including the influence of ISM, otherwise the strengths will further decrease), which are  within or approaching the observational windows of current magnetometer and EM detector based on atoms, superconducting quantum interference device, spin wave interferometer, coils-antennas and so on\cite{romalis,Taylor,Budker,Kominis,PhysRevLett.89.130801,Drung,Dang.Romalis.2010,Lee,Sheng,Groeger2006,Kitching,Ito,PhysRevApplied.6.064014,Breschi,Prance,Savukov,Granata,Grujic2015,Lucivero,PhysRevA.87.013413,Kulak,Balynsky,PhysRevLett.120.033401}.\\

\indent The strong magnetic fields of binary would be further significantly amplified by the amplification process, which had been widely studied\cite{0264-9381-16-6-201,Price719,PhysRevD.95.063016,PhysRevD.90.041502,PhysRevD.92.124034,PhysRevD.97.124039,PhysRevD.92.084064,2041-8205-769-2-L29,0264-9381-33-16-164001,0004-637X-809-1-39} as a key feature to understand the physical behaviours during the binary merger, and it perhaps lead to the strongest magnetic fields in the Universe\cite{Price719}.
Thus, the amplification process will not only result in the further stronger signals of the perturbed ULF-EMWs, but also lead to that the waveforms of perturbed ULF-EMWs have an enlargement and modification in corresponding time duration (of the amplification of magnetic fields).\\
\indent The particular polarizations of the perturbed ULF-EMWs caused by the tensorial and possible nontensorial polarizations of GWs will be a very special character. 
In frame of the GR, GWs have tensorial polarizations only ($\times$ and $+$ modes), but the generic metric theories predict up to six polarizations (including vector modes: $x$, $y$, and scalar modes: $b$, $\it{l}$)\cite{PhysRevLett.30.884,PhysRevD.8.3308}, and such additional polarizations relate to many important issues like the modified gravity and extra-dimensions of space.
Based on current researches\cite{PhysRevD.98.022008,Li.Wen.arXiv1712.00766}, specific relationship of how the polarizations of perturbed ULF-EMWs connect to the tensorial and nontensorial polarizations  of the GWs from binary mergers, will be addressed in this article, and typical examples are presented.\\
\indent The perturbed ULF-EMWs as a special type of EM counterparts of the GWs, sit in totally different frequency band to usual EM counterpart of GRBs, and moreover, the ULF-EMWs have accurate start time rather than the GRBs. The GRBs are usually assumed to be generated nearly at the same time to the GWs, but actually there would be still some unknown uncertainty of their start time [then lead to different arrival time (time lag or delay) to the Earth; e.g., for GW170817, it is $\sim$seconds, and for other cases, the time lag may be much longer, $\sim$hours or even  $\sim$days\cite{LIGOAA}, such as the supernova models\cite{Li_1998}]. Differently, for the perturbed ULF-EMWs, such uncertainty could be reduced, because under the frame of EM response to GWs, they just clearly have the same start time to the GWs from the source of binary, and thus may provide more accurate information for those analysis (relevant to many fundamental problems of gravity and cosmology, such as measurement of local Hubble constant, cosmological curvature, propagation speed of photon and graviton, extra-dimensions, Lorentz violation) based on the difference of  arrival times between the EM and GW signals.\\

\indent Plan of this article is as follows:
In Sect.\ref{estimation},  strengths of magnetic components of the perturbed ULF-EMWs caused by GWs of binary mergers are estimated. 
In Sect.\ref{section.polarizations},    particular polarizations of the perturbed ULF-EMWs depending on the tensorial and possible nontensorial polarizations of the GWs of binary mergers, are investigated. 
In Sect.\ref{section.waveforms},  the amplification process of magnetic fields and the modification of the perturbed ULF-EMWs are discussed.  
In Sect.\ref{summary},  summary and discussion  are given. \\

\section{Strengths of perturbed ultra-low-frequency EMWs caused by GWs from binary mergers}
\label{estimation}
\begin{figure}%
	\centering
	\subfigure[dipole surface magnetic fields]{%
		\label{}%
		\includegraphics[width=1.3 in]{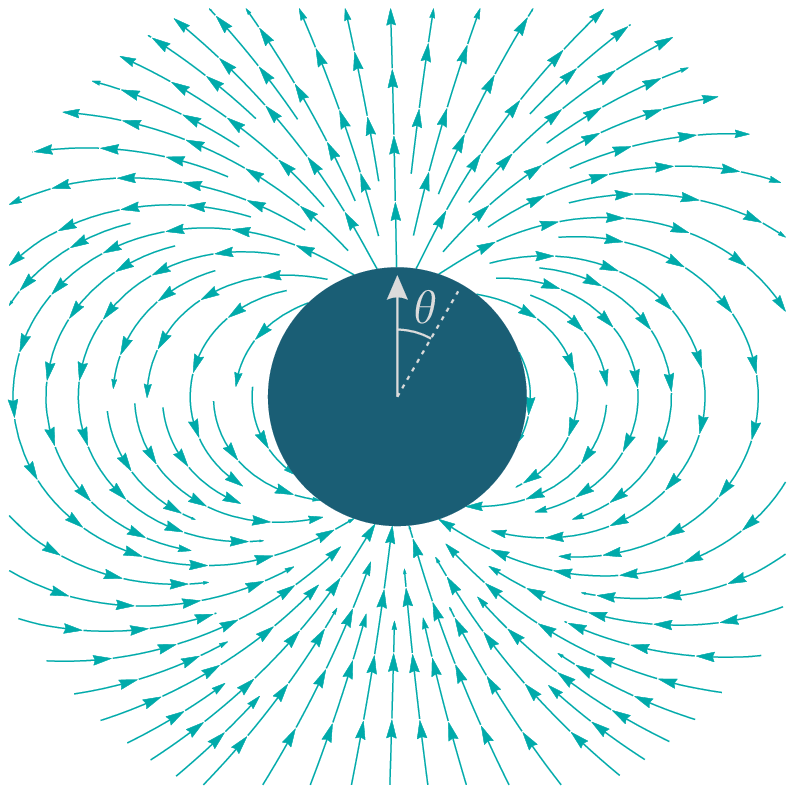}}
	\quad
	\subfigure[quadrupole surface magnetic fields]{%
		\label{}%
		\includegraphics[width=1.3 in]{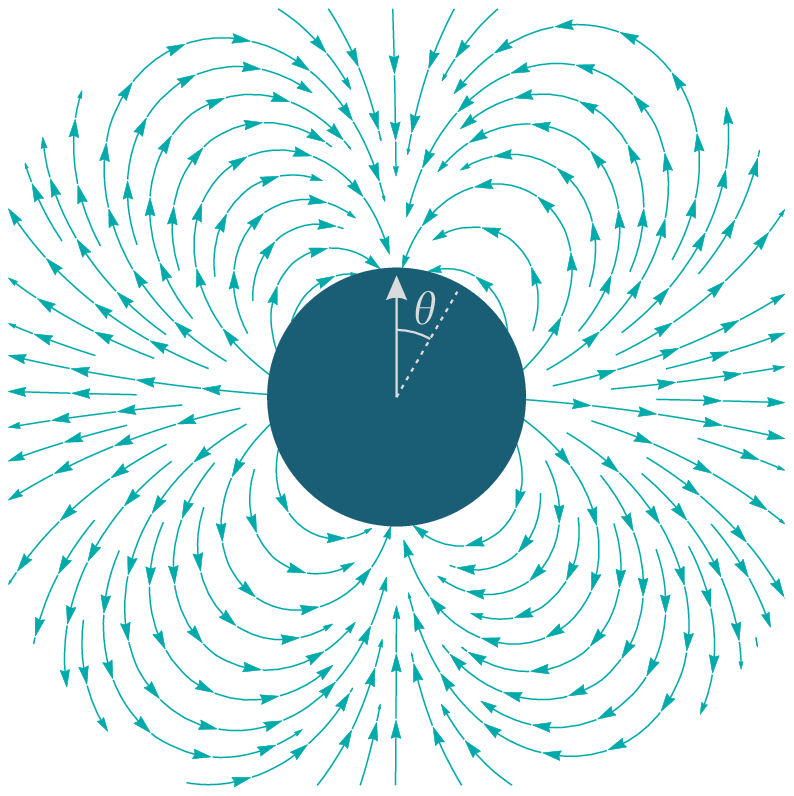}}

	\subfigure[decaying behaviors]{%
		\label{}%
		\includegraphics[width=3 in]{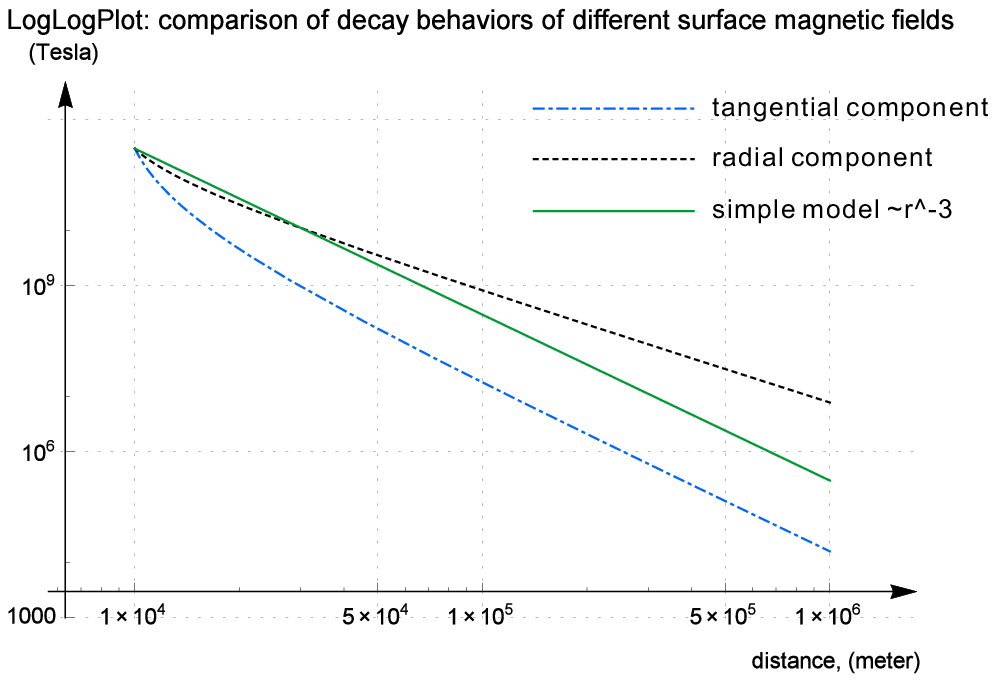}}
	\caption{a typical model of surface magnetic fields of neutron star employed for calculation. The (a) is dipole mode [Eq.(\ref{eq_Bsurf_dipole})] and (b) is quadrupole mode [Eq.(\ref{eq_Bsurf_quadrupole})]  of the surface magnetic fields. Comparison of decaying behaviours are shown in (c) among tangential and radial components of dipole mode, and simple model of magnetic fields $\sim r^{-3}$. The radius of neutron star is set as $10^4$m.}%
	\label{surfBfields}%
\end{figure}

In this section we estimate the strengths (at the Earth) of the perturbed ULF-EMWs  caused by interaction between the GWs of binary mergers and the ultra-strong magnetic fields of neutron star (or \mbox{magnetar}) of the same binary system.\\
\indent For the first step of estimation, instead of massive numerical computing, we apply a typical model\cite{PhysRevD.64.083008} of the surface magnetic fields of neutron stars for calculation, and it can be expressed as\cite{PhysRevD.64.083008}:
\begin{eqnarray}
\label{eqB}
&~&\textbf{B}^{surf}=\vec{\bigtriangledown}\times(\vec{r}\times\vec{\bigtriangledown}S),\nonumber\\
&~&S=S(l,m)=S_l^m(r)Y_l^m(\theta,\phi),\nonumber\\
&~&Y_l^m(\theta,\phi)=P_l^m(\cos\theta)e^{im\phi};
\end{eqnarray}
In spherical coordinates with  orthonormal basis of $\textbf{e}_r$, $\textbf{e}_{\theta}$ and $\textbf{e}_{\phi}$, the $\textbf{r}=r\textbf{e}_r$, and  $\textbf{B}=B_r\textbf{e}_r+B_{\theta}\textbf{e}_{\theta}+B_{\phi}\textbf{e}_{\phi}$.  The $S$ is expanded in a series of spherical harmonics, and $P_l^m(\cos\theta)$ is the Legendre polynomial. For $l=1,m=0$, it corresponds to
the dipole mode:
{\color{black}
\begin{eqnarray}
\label{eqS10}
S(1,0)&=&C\frac{\cos\theta}{r^2}\sum _{\nu =0}^{\infty }a_{\nu}(\frac{2M}{r})^{\nu},\nonumber\\
 a_{\nu}&=&3/(3+\nu),~(for~\nu=0,1,2,3,...).
\end{eqnarray}
}
From Eqs.(\ref{eqB}) to (\ref{eqS10}), the dipole component of surface magnetic field is:
\begin{eqnarray}
\label{eq_Bsurf}
&~&\textbf{B}^{surf}(1,0)=\vec{\bigtriangledown}\times( \vec{r}\times\vec{\bigtriangledown}S(1,0)  )\nonumber\\
&=&C_1\cos\theta\frac{1}{r^3}\sum _{\nu =0}^{\infty }a_{\nu}(\frac{2M}{r})^{\nu}\vec{e}_r\nonumber\\
&+&C_1\sin\theta\frac{1}{r^3h}\sum _{\nu =0}^{\infty }(\nu+1)a_{\nu}(\frac{2M}{r})^{\nu}\vec{e}_{\theta},
\end{eqnarray}
Sum the terms in Eq.(\ref{eq_Bsurf}), a typical form of neutron star surface magnetic field in dipole mode
can be obtained [see Fig.\ref{surfBfields} (a)]:
\begin{eqnarray}
\label{eq_Bsurf_dipole}
&~&\textbf{B}^{surf}_{di}(1,0)\nonumber\\
&=&2C_1\cos\theta\frac{1}{r^3}\frac{-3 r [r^2 \ln (1-\frac{2 M}{r})+2 M (M+r)]}{8M^3}\vec{e}_r \nonumber\\
&+&C_1\frac{\sin\theta}{r^3h}\frac{3 r^2 [2 M (\frac{M}{r-2 M}+1)+r \ln (1-\frac{2 M}{r})]}{4M^3}\vec{e}_{\theta}, \nonumber\\
\end{eqnarray}
{\color{black}
$\displaystyle M=\frac{G\tilde{m}}{c^2}$,  ($\tilde{m}$ is the mass of neutron star), and the metric $h$ is\cite{PhysRevD.64.083008}:
\begin{eqnarray}
\label{eqMandh}
~h=h(r)=(1-\frac{2G\hat m(r)}{rc^2})^{-\frac{1}{2}},
\end{eqnarray}
The $\hat m(r)$ is the   mass
function to determine the total mass enclosed within sphere of radius $r$.
When $r>$ neutron star radius, $\hat m(r)=\tilde{m}$ and then $\displaystyle\frac{2G\hat m(r)}{c^2}=2M$.
Therefore, the $h(r)$ very fast drops into 1 for $r>2M$, so in the calculation for the area $r>10$km (typical neutron star radius) we could approximately take the $h(r)$ as 1.\\
}

\indent Similarly,  for $l=2$, $m=0$, we have
the quadrupole mode of surface magnetic fields [see Fig.\ref{surfBfields} (b)]:
\begin{eqnarray}
\label{eq_Bsurf_quadrupole}
&~&\textbf{B}^{surf}_{quad}(2,0)
=3C_2(3\cos^2\theta-1)\frac{1}{r^4}\nonumber\\
&\cdot&\frac{-3 r [r^2 \ln (1-\frac{2 M}{r})+2 M (M+r)]}
{8M^3}\vec{e}_r\nonumber\\
&+&3C_2\cos\theta\sin\theta\frac{1}{r^4h}\nonumber\\
&\cdot&\frac{3 r [2 M (\frac{4 M^2}{r-2 M}+M+r)+r^2 \ln (1-\frac{2 M}{r})]}
{8M^3}\vec{e}_{\theta},\nonumber\\
\end{eqnarray}
\indent  Neutron star surface magnetic fields in  quadrupole mode would have comparable
strength to that in  dipole mode\cite{APJ.688.1258}, and the $C_1$ and $C_2$ are constants (with different dimensions) that have been
calibrated to satisfy the typical strength of surface magnetic fields (e.g. for magnetar, $\sim10^{11}$T). 
In dipole mode,
the tangential components [i.e. $\vec{e}_{\theta}$ component in Eq.(\ref{eq_Bsurf_dipole})]
have the maximum at polar angle $\theta=\pi/2$ [Fig.\ref{surfBfields}(a)],
and the radial components [i.e. $\vec{e}_{r}$ component in Eq.(\ref{eq_Bsurf_dipole})]
have the maximum around $\theta=0$ and $\pi$ (two poles).
Differently, in quadrupole mode[Eq.(\ref{eq_Bsurf_quadrupole})] the tangential components  have maximum around
$\theta=\pi/4$ and $3\pi/4$ [Fig.\ref{surfBfields}(b)], and the maximum of radial component is around $\theta=0, \pm\pi/2$, and $\pi$.\\
\indent We also compare above model with more simple models (i.e. just considering the magnetic fields decay  by $\sim r^{-3}$ or $\sim r^{-4}$), and we can see  that [Fig.\ref{surfBfields}(c)] the dipole magnetic fields decay faster than the simple model of $\sim r^{-3}$ in the near field close to the source. Actually, the decay behaviour of magnetic fields in near field area predominately impact the generation of the perturbed ULF-EMWs (we can see that in later part of this section), so we should use such typical model instead of the simple models to obtain more safe estimations.\\
\indent On the other hand, if the GWs of binary mergers contain possible nontensorial polarizations,  they can be generally express as:
\begin{eqnarray}\label{eq01}
&~&h_{\mu\nu} =\left( {{\begin{array}{*{20}c}
		0 &  0   & 0 &0\\
		0 & A_{+}+A_b&A_{\times}  &A_x\\
		0 & A_{\times}  &  -A_{+}+A_b& A_y\\
		0 &  A_x   & A_y& \sqrt{2}A_l \nonumber\\
		\end{array} }} \right)e^{i(\textbf{\textit{k}}_g\cdot\textbf{\textit{r}}-\omega t)},\\
\end{eqnarray}
the $+\&\times$, $x\&y$, $b\&l$ respectively represent the cross-\&plus- (tensor mode), $x$-\&$y$- (vector mode), $b$-\&$l$- (scalar mode) polarizations. Interaction of these GWs of binary mergers with the ultra-strong magnetic fields [background fields, Eqs. (\ref{eq_Bsurf_dipole}) and (\ref{eq_Bsurf_quadrupole})] of the neutron star of the binary system, will generate the perturbed ULF-EMWs, and such effect can be calculated by the electrodynamics equations in curved spacetime:
\begin{eqnarray}
\label{emeqcurved}
&~& \frac{1}{\sqrt{-g}}\frac{\partial}{\partial x^{\nu}}[\sqrt{-g}g^{\mu\alpha}g^{\nu\beta}(F_{\alpha\beta}^{(0)}+\tilde F_{\alpha\beta}^{(1)})]=\mu_0J^{\mu}, \nonumber\\
\label{eq18}
&~& \nabla_{\mu}F_{\nu\alpha}+\nabla_{\nu}F_{\alpha\mu}+\nabla_{\alpha}F_{\mu\nu}=0, \nonumber\\
\label{eq19}
&~& \nabla_{\alpha} F_{\mu\nu}=F_{\mu\nu,\alpha}-\Gamma^{\sigma}_{\mu\alpha}F_{\sigma\nu}-\Gamma^{\sigma}_{\nu\alpha}F_{\mu\sigma},
\end{eqnarray}
Due to previous works\cite{Li.Wen.arXiv1712.00766,FYLi_PRD80_2009,FYLi_EPJC_2008,PRD104025,WenEPJC2014,Boccaletti_NuovoCim70_1970},  the E and B components of the perturbed ULF-EMWs for an accumulation distance of $\Delta\textit{L}$ (small enough) were given:
\begin{eqnarray}
\label{eqEB}
\tilde{E}^{(1)}&=&A\hat{B}_{surf}^{(0)}k_g^{}c\Delta\textit{L}\exp[{i(\textbf{\textit{k}}_g\cdot\textbf{\textit{r}}-\omega t)}],\nonumber\\
\tilde{B}^{(1)}&=&A\hat{B}_{surf}^{(0)}k_g^{}\Delta\textit{L}\exp[{i(\textbf{\textit{k}}_g\cdot\textbf{\textit{r}}-\omega t)}],
\end{eqnarray}
here, ``$A$'' is the GW amplitude of tensorial modes ($A_{+}$, $A_{\times}$), or of nontensorial modes [here, only for ($A_x$, $A_y$), but not for ($A_b$, $A_l$), the reason is explained below]. The $\hat{B}_{surf}^{(0)}$ can be transverse magnetic fields [perpendicular to direction of GW propagation, e.g., the tangential components of Eqs. (\ref{eq_Bsurf_dipole}) and (\ref{eq_Bsurf_quadrupole})], or can be longitudinal magnetic fields [along the direction of GW propagation, e.g., the radial components of Eqs. (\ref{eq_Bsurf_dipole}) and (\ref{eq_Bsurf_quadrupole})]. \\
\indent Importantly, the tensorial GWs can interact with the transverse magnetic fields but cannot with the longitudinal magnetic fields, and contrarily, the nontensorial GWs can interact with the longitudinal magnetic fields but cannot with the transverse magnetic fields\cite{Li.Wen.arXiv1712.00766}.  Thus, in this article,  we only consider the vector modes of ($A_x$, $A_y$) for the nontensorial GWs, because the the longitudinal magnetic fields can only interact with ($A_x$, $A_y$) GWs and cannot interact with $A_b$ or $A_l$ GWs\cite{Li.Wen.arXiv1712.00766}.\\

\begin{figure}[!htbp]
	\centerline{\includegraphics[width=3.3in]{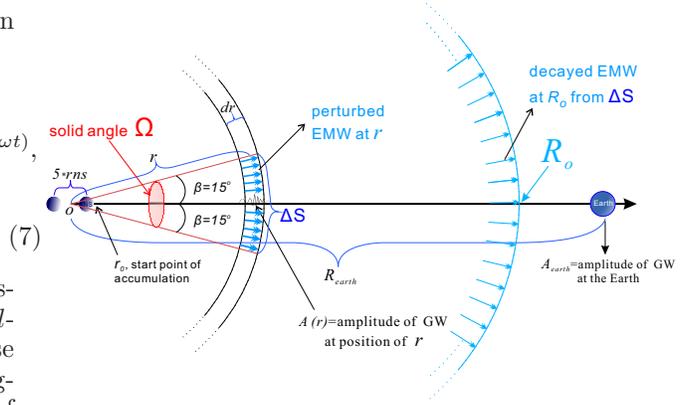}}
	\begin{spacing}{1.2}
		\caption{\footnotesize{\textbf{{\color{black}Scheme for the calculation for accumulated perturbed ULF-EMWs.}}
			 }}
		\label{int}
	\end{spacing}
\end{figure}

\indent To calculate the specific form of the perturbed ULF-EMWs is a much more complicated task, and it requires massive numerical computing for such non-stationary process. However, here, as the first step, at least we can analyze for an instantaneous state within a short duration, to make a brief and conservative estimation of the order of amplitude of the ULF-EMW strengths. Therefore, taking a typical example, we can consider an instantaneous situation in the late-inspiral phase shown in Fig. \ref{int}, where the binary evolution is very close to the merger time ($t=0$, defined as the time when the amplitude of GW reaches the maximum), e.g., only some  millisecond before the merger time, and the distance between the two centers of stars in the binary is set to $5\cdot rns$ ($rns=10^4$meter,  typical radius of neutron star). We can integrate the contributions [given by Eq. (\ref{eqEB}), replace the $\Delta L$ by the $dr$, and set the propagation factor as 1 to take the peak value] of generation of the perturbed ULF-EMWs of every small accumulation distance ``$dr$'', from the  $r_0$ (start point of the accumulation, set as $3.5\cdot rns$ here)  until some end point of accumulation at $R_{acc}$.\\
{\color{black}\indent In the  Fig. \ref{int},  we  calculate the contribution only within the solid angle $\Omega$, and in this  small $\Omega$, we consider the GWs to decay spherically, and approximately consider the magnetic fields of neutron star are isotropic [because even at the edge of $\Omega$ where $\beta=15^\circ$ (or $\pi/12$),  the magnetic fields have strength of $\cos({\pi}/{12})\times$maximum = 0.97 $\times$maximum $\approx$ 1$\times$maximum (the maximum appears at $\beta=0$)].
Therefore, at distance of $r$, we only include the contribution of perturbed ULF-EMWs generated in the region with thickness of ``$dr$'' and size of $\Delta S$ (the spherical cap to cover the solid angle $\Omega$). For very far observers, this disk-like source (on $\Delta S$) could be treated as a point source of EMWs with the total power concentrated into a point at the centre of $\Delta S$, and decay spherically (from this point) into far distance of $R_o$ (observer's distance). Thus, at the $R_o$, the energy flux density of the decayed EMWs is  
$\displaystyle\frac{total~power}{2\pi (R_o-r)^2}=\frac{c(\tilde{B}^{(1)}(R_o))^2}{\mu_0}$ [only consider the right half of the sphere, so the ``total power'' is divided by $2\pi (R_o-r)^2$], and because $\displaystyle\frac{total~power}{2\pi R_o^2}<\frac{total~power}{2\pi (R_o-r)^2}$, so we take the form of	 $\displaystyle\frac{total~power}{2\pi R_o^2}$ for a safe simplification.
Given that the ``total power'' of the perturbed ULF-EMWs (from $\Delta S$) is $\displaystyle\frac{c(\tilde{B}^{(1)}(r))^2}{\mu_0}\times\Delta S=\frac{c(\tilde{B}^{(1)}(r))^2}{\mu_0}\times\Omega r^2$, then we have
$\displaystyle\tilde{B}^{(1)}(R_o)=\tilde{B}^{(1)}(r)\frac{r}{R_o}\sqrt{\frac{\Omega}{2\pi}}$. 
For conciseness, let $\displaystyle\xi=\sqrt{\frac{\Omega}{2\pi}}$. Therefore, there are terms of ``$\displaystyle\frac{r}{R_o}$'' and ``$\xi$'' in the integral of Eq. (\ref{B_dipole_tensor}) [see below, and similarly for Eqs. (\ref{B_dipole_nontensor}) to (\ref{perturbedBfield})],  and if take the $\displaystyle\beta=\frac{\pi}{12}$, this factor $\xi$ is about 0.18.}

\indent Actually, the accumulation will continue after the $R_{acc}$, but we drop such part because it is decaying fast, and here we set the $R_{acc}$  into   only $10^5$meter for a safe estimation.
Therefore, together with Eqs. (\ref{eq_Bsurf_dipole}), (\ref{eq_Bsurf_quadrupole}) and (\ref{eqEB}), we work out the accumulated perturbed ULF-EMWs caused by the tensorial GWs interacting with transverse surface magnetic field  (tangential, or $\vec{e_{\theta}}$ component) of the dipole mode, and the strengths of their magnetic components has the form: 
\begin{widetext}
{\color{black}
\begin{eqnarray}
\label{B_dipole_tensor}
&~&\tilde{B}_{prtbd-tsr}^{dipole}=\int^{R_{acc}}_{r_0}(\frac{R_{earth}A_{earth}}{r})  \frac{\sin\theta C_1}{r'^3h}\frac{3 r'^2 [2 M (\frac{M}{r'-2 M}+1)+r'\ln(1-\frac{2 M}{r'})]}{4M^3}   \frac{\omega}{c}\frac{r}{R_o}\xi dr \nonumber\\
&=& \xi\frac{ \sin\theta 3 C_1 A_{earth} R_{earth} \omega}{4c h M^3 R_o}[R_{acc}'\ln(1-\frac{2M}{R_{acc}'})-r_0'\ln(1-\frac{2M}{r_0'})+M \ln \frac{R_{acc}'({r_0'}-2 M)}{{r_0'} (R_{acc}'-2 M)}], 
\end{eqnarray}
}
\end{widetext}
The term ($\displaystyle\frac{R_{earth}A_{earth}}{r}$) represents $A(r)$ (amplitude of GW at position of $r$), where the $R_{earth}$ and $A_{earth}$ are the distance of binary to the Earth and the amplitude of GW at the Earth. The subscript ``$prtbd$'' and ``$tsr$'' of above $\tilde{B}_{prtbd-tsr}^{dipole}$ mean ``perturbed EMWs'' and ``caused by tensorial GWs''; the superscript ``$dipole$'' means here we include the dipole mode of surface magnetic fields for calculation. The  $\omega$ is the angular frequency. Besides, here we use the ``$r'$'' for the part of magnetic field in the Eq.(\ref{B_dipole_tensor}),  due to that the center of the neutron star is not at the center of the binary (where $r=0$), but at the $r=2.5*rns$ (also see Fig. \ref{int}) for our calculation, i.e., the radial coordinate of neutron star (included for calculation) is shifted into $r'=r-2.5*rns$; also, we note the $R_{acc}'=R_{acc}-2.5rns$, and $r_0'=r_0-2.5rns$, respectively. 
When $R_o > R_{acc}$, as also mentioned above, we only include the contribution of perturbed ULF-EMWs before the $R_{acc}$, and when  $R_o\leqslant R_{acc}$, we set $R_{acc}\rightarrow R_o$ (replace $R_{acc}$ by $R_o$). \\

\indent  In a similar way, the strength of magnetic component of the perturbed ULF-EMWs caused by nontensorial GWs interacting with longitudinal magnetic field  (radial, or $\vec{e_r}$ component) of the dipole mode, can be obtained:
\begin{widetext}
{\color{black}
\begin{eqnarray}
\label{B_dipole_nontensor}
&~&\tilde{B}_{prtbd-ntsr}^{dipole}=\int^{R_{acc}}_{r_0}(\frac{R_{earth}A_{earth}}{r}) 2C_1\cos\theta\frac{1}{r'^3}\frac{-3 r' [r'^2 \ln (1-\frac{2 M}{r'})+2 M (M+r')]}{8M^3}  \frac{\omega}{c}\frac{r}{R_o} \xi dr \nonumber\\
&=&\xi\frac{\cos \theta 3 C_1 A_{earth} R_{earth} \omega  }{4 c M^3  R_o} [2 M^2 \frac{r_0'-R_{acc}'}{r_0'R_{acc}'} +2 M \ln \frac{r_0' (R_{acc}'-2 M)}{R_{acc}' (r_0'-2 M)}+ r_0'\ln \frac{r_0'-2 M}{r_0'}-R_{acc}' \ln \frac{R_{acc}'-2 M}{R_{acc}'}],~~~~~~
\end{eqnarray}
}
\end{widetext}
The subscript ``$ntsr$'' of above $\tilde{B}_{prtbd-ntsr}^{dipole}$ means ``caused by nontensorial GWs''. The same,  the magnetic component of the perturbed ULF-EMWs caused by tensorial GWs interacting with transverse magnetic fields  ($\vec{e_{\theta}}$ component)  of the quadrupole mode, has the form:
\begin{widetext}
{\color{black}
\begin{eqnarray}
\label{B_quad_tensor}
&~&\tilde{B}_{prtbd-tsr}^{quad.}=\int^{R_{acc}}_{r_0}(\frac{R_{earth}A_{earth}}{r}) 3C_2\cos\theta\sin\theta\frac{1}{r'^4h}\frac{3 r' [2 M (\frac{4 M^2}{r'-2 M}+M+r')+r'^2 \ln (1-\frac{2 M}{r'})]}{8M^3} \frac{\omega}{c}\frac{r}{R_o} \xi dr \nonumber\\
&=&\xi\frac{ 9\sin \theta\cos \theta  C_2  R_{earth}A_{earth} \omega  }{ 8 c h M^3 R_o } [ M^2 (\frac{1}{R_{acc}'^2}-\frac{1}{r_0'^2})+\ln \frac{r_0' (R_{acc}'-2 M)}{R_{acc}' (r_0'-2 M)}+\sum_{k=1}^{\infty }\frac{(  2M/R_{acc}')^k}{k^2}  +\sum_{k=1}^{\infty }\frac{(  2M/r_0')^k}{k^2}],\nonumber\\
\end{eqnarray}
}
\end{widetext}
The superscript ``$quad.$ ''of above $\tilde{B}_{prtbd-tsr}^{quad.}$ means the quadrupole mode of magnetic fields are included for calculation.
Further,  the magnetic component of  the   perturbed ULF-EMWs caused by nontensorial GWs interacting with longitudinal magnetic field ($\vec{e_{r}}$ component) of the quadrupole mode, can be given:
\begin{widetext}
{\color{black}
\begin{eqnarray}
\label{B_quad_nontensor}
&~&\tilde{B}_{prtbd-ntsr}^{quad.}=\int^{R_{acc}}_{r_0}(\frac{R_{earth}A_{earth}}{r}) 3C_2(3\cos^2\theta-1)\frac{1}{r'^4}\frac{-3 r' [r'^2 \ln (1-\frac{2 M}{r'})+2 M (M+r')]}
{8M^3} \frac{\omega}{c}\frac{r}{R_o}\xi dr \nonumber\\
&=&\xi\frac{ 9(3\cos^2\theta-1) C_2 R_{earth}A_{earth}  \omega  }{ 8 c  M^3 R_o } [(\frac{M(M+2R_{acc}')}{R_{acc}'^2}-\frac{M(M+2r_0')}{r_0'^2}-\sum_{k=1}^{\infty }\frac{(  2M/R'_{acc})^k}{k^2}+\sum_{k=1}^{\infty }\frac{(  2M/r_0')^k}{k^2}],~~~~~~
\end{eqnarray}
}
\end{widetext} 
If using simple models of the surface magnetic fields of  neutron stars, which just decay by $\sim r^{-n}$(n=3, 4, ...), the strengths of magnetic components of the perturbed ULF-EMWs are:
{\color{black}
\begin{eqnarray}
\label{perturbedBfield}
&&\tilde{B}_{prtbd}^{simple}=\int^{R_{acc}}_{r_0}(\frac{R_{earth}A_{earth}}{r})(\frac{B_0\cdot rns^n}{r'^n})\frac{\omega}{c}\frac{r}{R_o}\xi dr \nonumber\\
&=&\xi\frac{R_{earth}A_{earth}rns^nB_0\omega}{(n-1)cR_o}(\frac{1}{r_0'^{n-1}}-\frac{1}{R_{acc}'^{n-1}})
\end{eqnarray}
}
\begin{table*}[!htbp]
{\color{black}
	\caption{\label{perturbedBfarfield}
		Strengths (in Tesla) of magnetic components  of the perturbed ULF-EMWs at the Earth [based on Eqs. (\ref{B_dipole_tensor}) to (\ref{perturbedBfield})]. 
		The cell of ``dipole B nontensorial GWs''  represents cases that the dipole mode of magnetic fields of neutron stars and the nontensorial GWs of binary mergers are included for calculation, and similarly for other cells.
		Here, the cases with surface ``magnetic fields of neutron star'' $=1.0\times10^{11}$, $1.0\times10^{12}$ and $1.0\times10^{8}$Tesla respectively represent the magnetar cases, the cases considering the amplification of binary magnetic fields, and the cases of only normal neutron stars.  	
		The frequency here is set to fISCO (smaller than the maximum frequency of the merger).	
		The strengths of such signals (also special EM counterparts of the GWs of binary mergers)  are  {\color{black}  mainly around  $\sim10^{-12}$Tesla to $\sim10^{-17}$Tesla}, and for the magnetar cases it could reach $\sim10^{-12}$ to $\sim10^{-13}$Tesla.		   
	}	
	\begin{tabular}{ccccccccc}
		\hline
		\hline
		GW	&distance&magnetic&\multicolumn{6}{c}{magnetic component (Tesla) of perturbed ULF-EMWs around the Earth}\\
		\cline{4-9}
		amplitude	&of binary&fields of&dipole B&dipole B&quadrupole B&quadrupole B&simple&simple\\
		around	&sources& neutron &tensorial&nontensorial&tensorial&nontensorial&model&model\\
		Earth	&to Earth&~star(Tesla) &GWs &GWs&GWs&GWs&$B\sim r^{-4}$&$B\sim r^{-3}$\\
		\hline
		~\\	
		
	&	&$1.0\times10^{11}$&$8.9\times10^{-13}$&$1.6\times10^{-12}$&$8.1\times10^{-13}$&$1.2\times10^{-12}$&$2.0\times10^{-12}$&$3.0\times10^{-12}$\\
&$40Mpc$&$1.0\times10^{12}$&$8.9\times10^{-12}$&$1.6\times10^{-11}$&$8.1\times10^{-12}$&$1.2\times10^{-11}$&$2.0\times10^{-11}$&$3.0\times10^{-11}$\\
&&$1.0\times10^{8}$&$8.9\times10^{-16}$&$1.6\times10^{-15}$&$8.1\times10^{-16}$&$1.2\times10^{-15}$&$2.0\times10^{-15}$&$3.0\times10^{-15}$\\
	
		$10^{-21}$	&&&&&&&&\\
	&	&	$1.0\times10^{11}$&$8.9\times10^{-13}$&$1.6\times10^{-12}$&$8.1\times10^{-13}$&$1.2\times10^{-12}$&$2.0\times10^{-12}$&$3.0\times10^{-12}$\\
&$4Mpc$&$1.0\times10^{12}$&$8.9\times10^{-12}$&$1.6\times10^{-11}$&$8.1\times10^{-12}$&$1.2\times10^{-11}$&$2.0\times10^{-11}$&$3.0\times10^{-11}$\\
&&$1.0\times10^{8}$&$8.9\times10^{-16}$&$1.6\times10^{-15}$&$8.1\times10^{-16}$&$1.2\times10^{-15}$&$2.0\times10^{-15}$&$3.0\times10^{-15}$\\
		~\\		
	&	&$1.0\times10^{11}$&$8.9\times10^{-14}$&$1.6\times10^{-13}$&$8.1\times10^{-14}$&$1.2\times10^{-13}$&$2.0\times10^{-13}$&$3.0\times10^{-13}$\\	
&$40Mpc$&$1.0\times10^{12}$&$8.9\times10^{-13}$&$1.6\times10^{-12}$&$8.1\times10^{-13}$&$1.2\times10^{-12}$&$2.0\times10^{-12}$&$3.0\times10^{-12}$\\
&&$1.0\times10^{8}$&$8.9\times10^{-17}$&$1.6\times10^{-16}$&$8.1\times10^{-17}$&$1.2\times10^{-16}$&$2.0\times10^{-16}$&$3.0\times10^{-16}$\\
		
		$10^{-22}$	&&&&&&&&\\
		&	&$1.0\times10^{11}$&$8.9\times10^{-14}$&$1.6\times10^{-13}$&$8.1\times10^{-14}$&$1.2\times10^{-13}$&$2.0\times10^{-13}$&$3.0\times10^{-13}$\\
		&$400Mpc$&$1.0\times10^{12}$&$8.9\times10^{-13}$&$1.6\times10^{-12}$&$8.1\times10^{-13}$&$1.2\times10^{-12}$&$2.0\times10^{-12}$&$3.0\times10^{-12}$\\
		&&$1.0\times10^{8}$&$8.9\times10^{-17}$&$1.6\times10^{-16}$&$8.1\times10^{-17}$&$1.2\times10^{-16}$&$2.0\times10^{-16}$&$3.0\times10^{-16}$\\
		~\\		
		\hline
	\end{tabular}
}
\end{table*}

\begin{figure}[!htbp]
	\centerline{\includegraphics[width=3.3in]{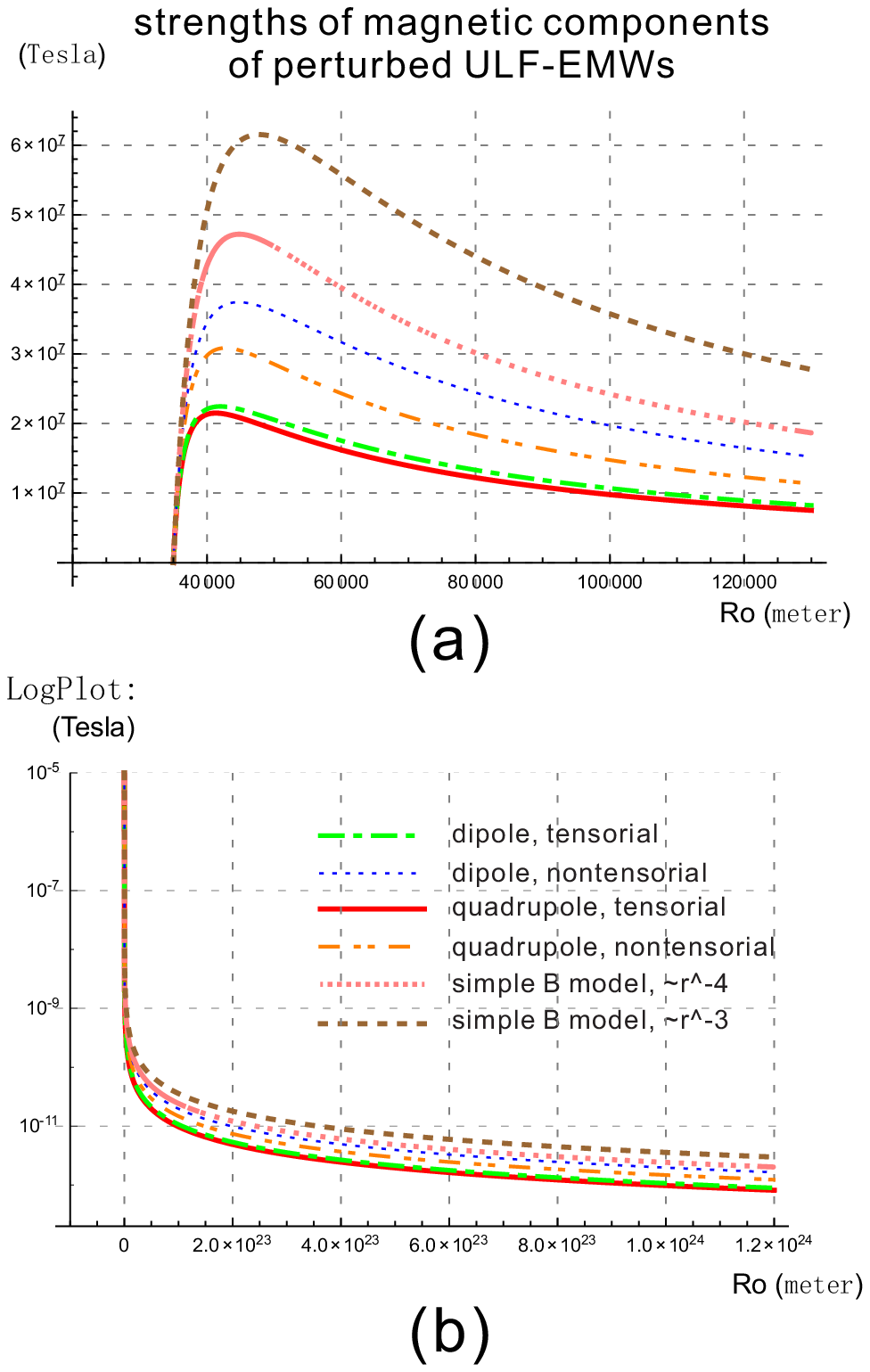}}
	\begin{spacing}{1.2}
\caption{\footnotesize{\textbf{{\color{black}			Strengths of magnetic components of perturbed ULF-EMWs caused by interaction between GWs from binary mergers and ultra-strong magnetic fields of neutron star in the binary system.} }
			The curves are from Eqs. (\ref{B_dipole_tensor}) to (\ref{perturbedBfield}), and the case of ``dipole, tensorial'' means the dipole mode of surface magnetic fields of neutron stars and the tensorial GWs of binary mergers are included for calculation, similarly for other cases. Subfigure (a) shows examples of curves of the strengths for range from the source to region near the $R_{acc}$ (end point of accumulation we include for calculation).  Subfigure (b) shows curves for the range from the $R_{acc}$ to  very far observation point $R_o=40Mpc$.  Here, the surface magnetic field of neutron star is set to $1.0\times10^{11}T$ (for magnetar cases), and the dimensionless amplitude of GWs at the Earth is set to $1.0\times10^{-21}$. }}
		\label{Bcurves}
	
	\end{spacing}
\end{figure}

\indent Above  Eqs. (\ref{B_dipole_tensor}) to (\ref{perturbedBfield}) estimate the strengths (of magnetic components) of the perturbed ULF-EMWs for various cases: dipole-tensorial, dipole-nontensorial, quadrupole-tensorial, quadrupole-nontensorial and simple models of magnetic fields; these results do not contain information of propagation factors or specific waveforms, and only provide estimations for the levels of strengths.  Table \ref{perturbedBfarfield}  and Fig. \ref{Bcurves} show examples  of these strengths for typical parameters. \\ 

\indent In Table \ref{perturbedBfarfield} we can find that the levels of magnetic components of the perturbed ULF-EMWs are {\color{black} mainly about $\sim10^{-12}$Tesla to  $\sim10^{-17}$Tesla}  at the Earth, for three type of cases: magnetar case (surface magnetic field $\sim10^{11}$Tesla), amplification case (surface magnetic field $\sim10^{12}$Tesla), and normal neutron star case (surface magnetic field $\sim10^{8}$Tesla). \\

\indent Also, in Fig. \ref{Bcurves} we find that the behaviours of signals among various cases are generally consistent, and the strengths based on simple models of magnetic fields ($\sim r^{-3}$, $\sim r^{-4}$) are generally larger than the strengths based on the dipole or quadrupole magnetic fields, which drop more fast in the near field. 
Actually, even for cases with just normal neutron stars (not magnetars), the magnetic fields of the binary would be greatly amplified by 3 (or more) orders of magnitude and easily reaching $10^{16}$G ($10^{12}$Tesla) or higher by the process of magnetic fields amplification (see Sect. \ref{section.waveforms}). Besides, results in Table \ref{perturbedBfarfield} indicate that the distance does not impact the level (at Earth) of the perturbed ULF-EMWs given the same GW amplitudes at the Earth, because larger distance of sources (binary) requires higher GW amplitudes at the sources, and then it leads to stronger generation of perturbed ULF-EMWs which whereas also need to decay for longer distance to the Earth, so their effects offset each other and thus compositively result in the irrelevance to the distance.\\

\indent The example case of Fig. \ref{int} (which above estimations based on) is for the pre-merger (late-inspiral) phase. However,  even for much more complicated merger and post-merger phases, the amplitudes of GWs are normally in the same order of magnitude to the GWs in the late-inspiral phase (at least within many milliseconds after the merger time), and meanwhile the magnetic fields would also be amplified into level of $\sim 10^{12}$Tesla or even higher (see Sections below), and thus, the perturbed ULF-EMWs (in the merger and post-merger phases) related to  levels of both GWs and  magnetic fields, should have stronger strengths than the values estimated above for the case of Fig. \ref{int}  for the pre-merger (late-inspiral) phases. Therefore, although here the detailed calculations for merger/post-merger phases are not given (need massive numerical computing, will be carried out in subsequent other works), the above estimations should also be safe for these phases.\\


\indent Another important issues is the propagation of the perturbed ULF-EMWs via the plasma medium, including the Earth ionosphere and the interstellar medium (ISM). The frequencies of the EMWs need to be higher than the plasma frequency (or they cannot propagate through the plasma), and the plasma frequency is depending on the electron density of the plasma, in the following way\cite{Gurnett1489,Gurnett2005}:\
\begin{eqnarray}
\label{plasmaFrequency}
\text{plasma\;frequency} \approx  8.97\;\text{kHz}\;(\frac{n_e}{1\cdot cm^{-3}})^{\frac{1}{2}}
\end{eqnarray}
where $n_e$ is the electron density (number density, per $cm^3$). The $n_e$ of plasma in the Earth ionosphere is quite high and thus the ULF-EMWs cannot propagate through it, unless there are some leakage. Therefore, it is much more possible to capture the ULF-EMWs by ULF detectors and magnetometers placed in the spacecrafts around the Earth. Whereas, on the other hand, we also need to consider the possible influence by the interstellar medium (ISM). In the region outside the Milky way, whether the ULF-EMWs will be influenced by the ISM, is depending on the position of the source of binary, but at least, we must take account of the influence of ISM within the Milky way. The electron density normally declines in perpendicular direction from the midplane of the Milky way to the outer region, and for the location of the Sun (8.5kpc from the Galactic Center), the $n_e$  is about $\sim 0.04 \text{cm}^{-3}$ (corresponding to plasma frequency of about \mbox{1.8 kHz}) around the Sun, and the  $n_e$  declines with increasing distance from the midplane, e.g., into value of $\sim 0.002 \text{cm}^{-3}$  (corresponding to plasma frequency of about 0.4 kHz) at distance of 2 kpc\cite{draine2010physics}. \\
\indent Thus,   we should mainly focus on the perturbed ULF-EMWs (or their EM components) with frequencies higher than \mbox{1.8 kHz}, as the more possible observational targets. Therefore, we discuss two stages [(1) merger and post-merger stage, (2) late-inspiral stage] of the binary merger that would produce the GWs in the band above \mbox{1.8 kHz} (then lead to the perturbed ULF-EMWs in such frequencies).\\
\indent (1) For the merger and post-merger stage, the GWs can easily surpass \mbox{1.8 kHz} into several kHz\cite{Cao2019IJMPD,Shibata10.1143,PhysRevD.80.064037,PhysRevLett.107.051102}, e.g. the fundamental oscillations, quasiperiodic waves,  formed transiently after the onset of merger (frequencies can be 2 to 3 kHz\cite{Cao2019IJMPD,PhysRevLett.107.051102,Shibata10.1143,PhysRevD.80.064037,PhysRevD.61.064001,PhysRevD.65.103005,PhysRevD.50.6247,Faber2012}). Crucially, in the merge and post-merge stage, the magnetic fields can be amplified into extremely high level of even $10^{12}$Tesla (Sect \ref{section.waveforms}), and thus, such GWs in several kHz band would produce  the perturbed ULF-EMWs in considerable strengths. 
Therefore, these perturbed ULF-EMWs produced in the merger and post-merger stage, can be above the plasma frequency of the ISM (as mentioned above, for position of solar system in the Milky way, around 1.8kHz).\\
\indent (2) For the late-inspiral stage, the binaries are usually investigated by quasiequilibrium configurations until their separations reach comparable size of the stars themselves, and the GWs frequencies  depend on parameters such as the mass of neutron star, orbital velocity,  separation and the compactness. E.g., the frequency of the pre-merger GW can be estimated by: $\displaystyle  f_{GW}\approx960 \text{Hz} (\frac{v^2}{0.12})^{3/2}(\frac{2.8M_{\odot}}{M_t})$ (where the $M_t$ is the total mass of binary; $v$ is the orbital velocity that can be set in 0.05 to 0.155\cite{PhysRevD.64.104017}). Thus, given a typical neutron star mass of 1.4 $M_{\odot}$, the GW frequency is in the range of 250 to 1350  Hz\cite{PhysRevD.64.104017}. Or, by use of a typical formula of the $f_{ISCO}=4396/(M/M_{\odot})$Hz\cite{Abadie_2010} (frequency of innermost stable circular orbit), for mass of NS as 1.4 $M_{\odot}$, the $f_{ISCO}=1570$Hz, and this agrees well to the above estimated value of $1350$ Hz. However, the low mass neutron stars are  still not theoretically ruled out [the theoretical bounds on the mass of neutron star and the mass ratio of  binary could  be possibly in large range for various equations of state (EOSs)\cite{PhysRevD.92.124007}]. E.g., with mass of 1.2 $M_{\odot}$, the frequencies of the  $f_{ISCO}=4396/2.4=1831$ Hz, which can be higher than the plasma frequency of \mbox{1.8 kHz} (for the very late inspiral phase similar to the situation in Fig. \ref{int}). Of course, for the lower mass neutron stars, the amplitude of the produced GWs would be  less  than the case with mass of around 1.3 to 1.4$M_{\odot}$ (could treat the amplitude is generally $\sim M^{\frac{5}{3}}$). 
E.g.,  if the GW amplitude decrease by 1 order of magnitude (namely, lead to that the perturbed ULF-EMWs also decrease by 1 order of magnitude), for the magnetar case (Table \ref{perturbedBfarfield}), i.e., the magnetic component of the ULF-EMWs will be depressed into level of $\sim10^{-13}$ to $\sim10^{-14}$Tesla.\\
\indent Therefore, for both the merger/post-merger stage and the pre-merger stage, the generation and propagation (to the Earth, above the plasma frequency of ISM) are possible in some parameter space. \\

\indent Although the propagation of the perturbed ULF-EMWs in the ISM is discussed above, again, here we do not include other effects of influence of the ISM, such as the dispersion, scattering, etc. Such specific experimental issues for practical astrophysical observations (could be detailedly addressed in other works) are much more complicated and should not be the topic focused in this article of only theoretical proposal with brief and preliminary estimations.

\section{Particular polarizations of the perturbed ULF-EMWs depending on tensorial and possible nontensorial polarizations of GWs from binary mergers}
\label{section.polarizations}
\begin{figure*}%
	\centering
	\subfigure[0$|$1$|$0$|$0$|$0$|$0]{%
		\label{}%
		\includegraphics[width=0.8 in]{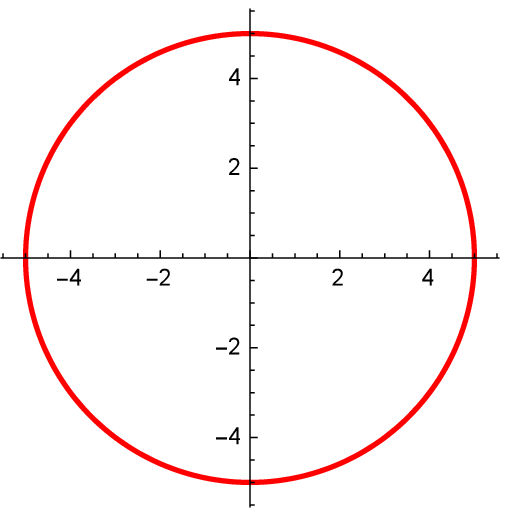}}
	\quad
	\subfigure[0$|$1$|$0.3$|$1$|$1$|$1]{%
		\label{}%
		\includegraphics[width=0.8 in]{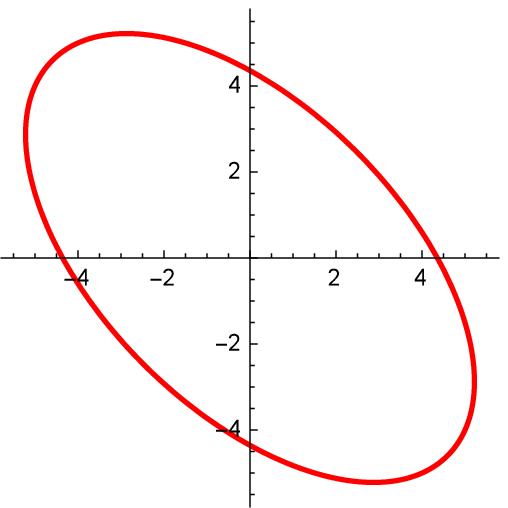}}
	\quad
	\subfigure[0$|$-1$|$0.7$|$1$|$1$|$1]{%
		\label{}%
		\includegraphics[width=0.8 in]{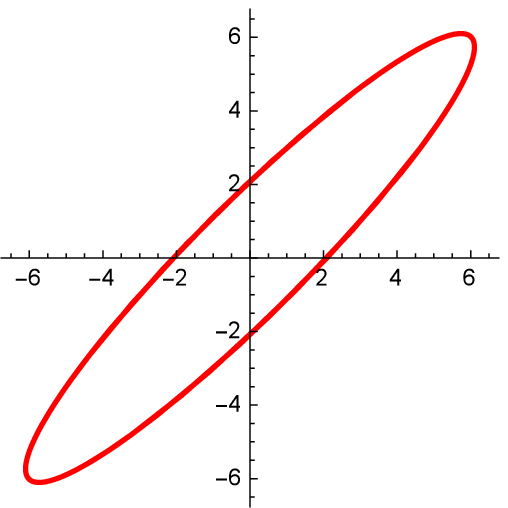}}
	\quad
	\subfigure[0$|$1$|$1$|$1$|$1$|$1]{%
		\label{}%
		\includegraphics[width=0.8 in]{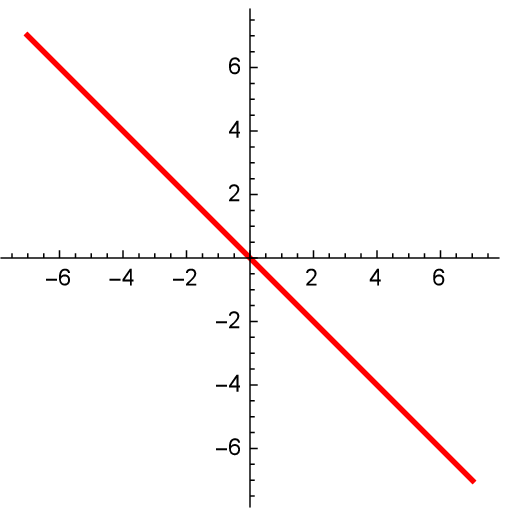}}
	\quad
	\subfigure[0$|$1$|$1.2$|$1$|$1$|$1]{%
		\label{}%
		\includegraphics[width=0.8 in]{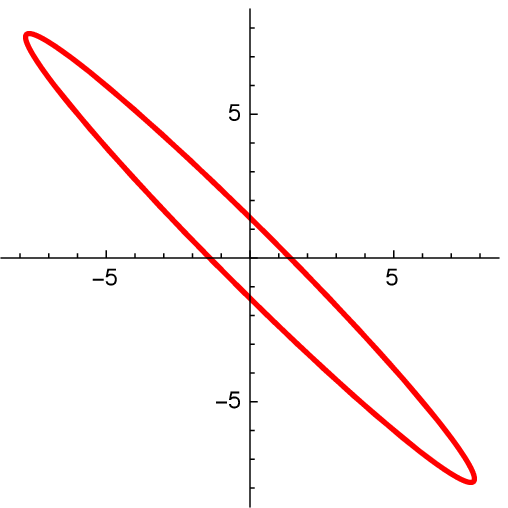}}\\
	\subfigure[$\frac{\pi}{2}|$1$|$1$|$1$|$1$|$1]{%
		\label{}%
		\includegraphics[width=0.8 in]{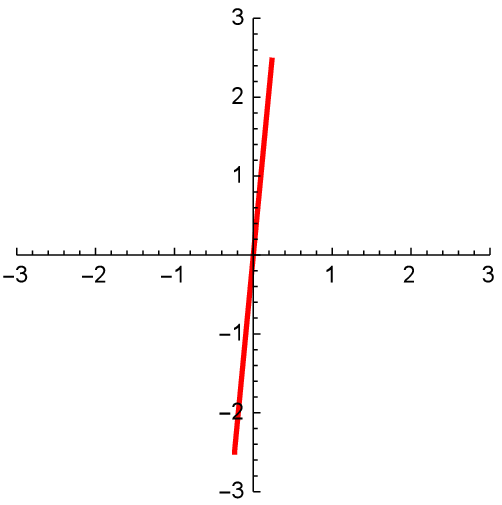}}
	\quad
	\subfigure[$\frac{\pi}{2}|$1$|$1$|$0$|$1$|$1]{%
		\label{}%
		\includegraphics[width=0.8 in]{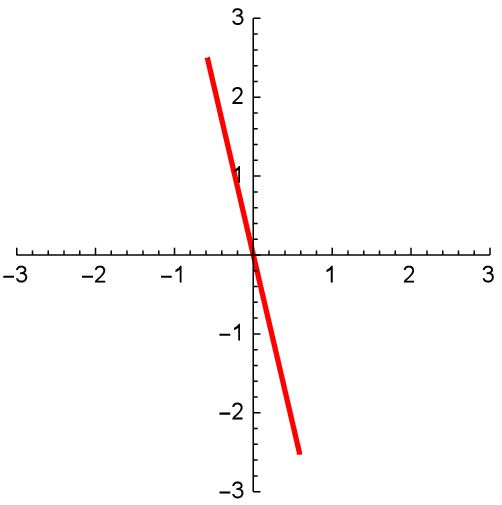}}
	\quad
	\subfigure[$\frac{\pi}{2}|$1$|$0.3$|$0.7$|$1$|$1.5]{%
		\label{}%
		\includegraphics[width=0.8 in]{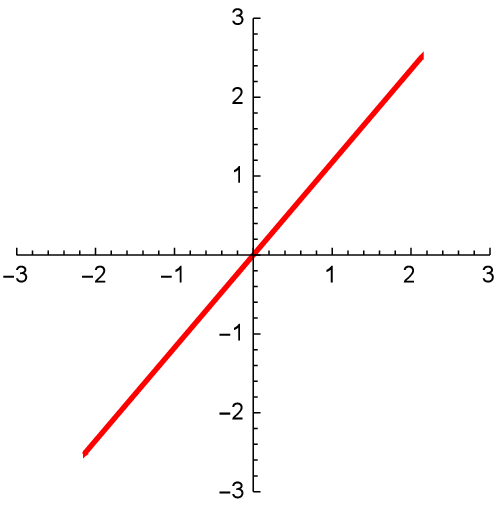}}
	\quad
	\subfigure[$\frac{\pi}{2}|$1$|$0$|$0$|$1$|$1]{%
		\label{}%
		\includegraphics[width=0.8 in]{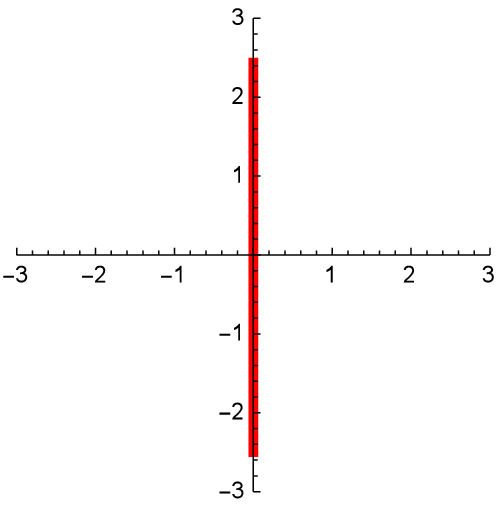}}
	\quad
	\subfigure[$\frac{\pi}{2}|$0$|$1$|$0$|$1$|$1]{%
		\label{}%
		\includegraphics[width=0.8 in]{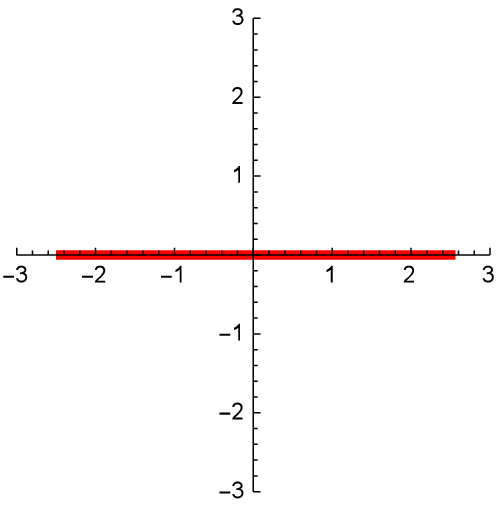}}\\
	\subfigure[$\frac{2\pi}{3}|$1$|$1$|$1$|$1$|$1]{%
		\label{}%
		\includegraphics[width=0.8 in]{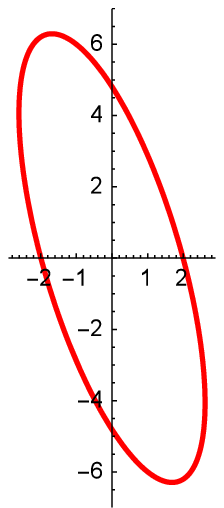}}
	\quad
	\subfigure[$\frac{2\pi}{3}|$1$|$1$|$0.5$|$0.1$|$0.1]{%
		\label{}%
		\includegraphics[width=0.8 in]{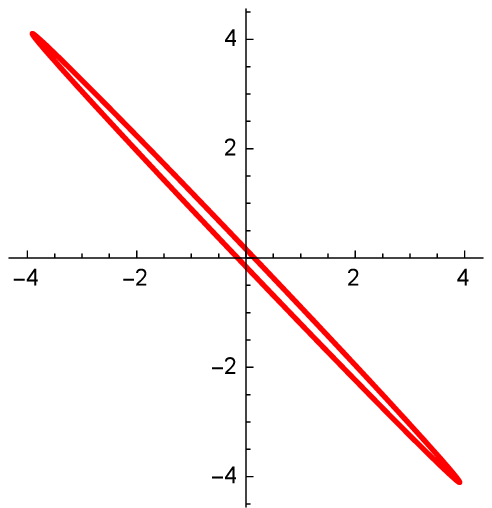}}
	\quad
	\subfigure[$\frac{2\pi}{3}|$0$|$0$|$1$|$0.3$|$0.5]{%
		\label{}%
		\includegraphics[width=0.8 in]{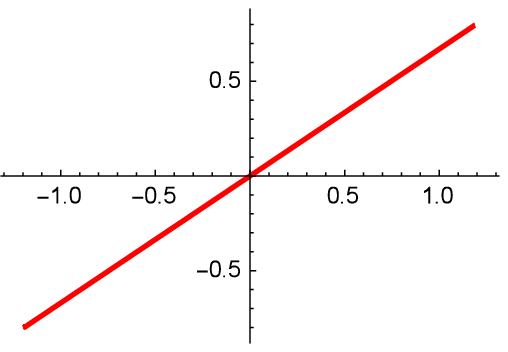}}
	\quad
	\subfigure[$\frac{2\pi}{3}|$1$|$0$|$0.1$|$1$|$1]{%
		\label{}%
		\includegraphics[width=0.8 in]{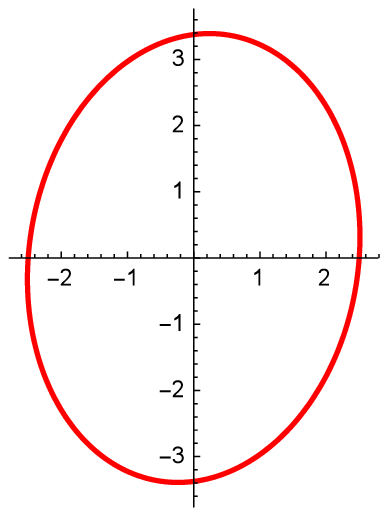}}
	\quad
	\subfigure[$\frac{2\pi}{3}|$1$|$0$|$0$|$0$|$0]{%
		\label{}%
		\includegraphics[width=0.8 in]{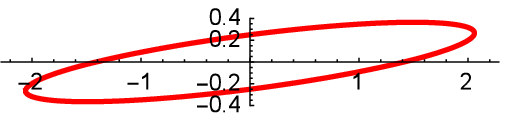}}\\
	\caption{Examples of  polarizations of perturbed ULF-EMWs caused by binary merger GWs having both tensorial and nontensorial modes by different focused parameters (influence of other parameters are ignored here) based on the connection relationship of Eq. (\ref{polarization4}) in Sect. \ref{section.polarizations}. The parameters (inclination angle, background magnetic fields, relative amplitudes of nontensorial GWs in vector modes) of every subfigure specifically means: $\iota|B_x^{(0)}|B_y^{(0)}|B_z^{(0)}|A_{V_x}|A_{V_y}$.}%
	\label{polarizations}%
\end{figure*}
\indent Some researches have been carried out\cite{PhysRevD.98.022008,PRL120.031104,arXiv1710.03794} for additional polarizations relevant to observations by LIGO. However, more information of specific properties of possible nontensorial GWs from binary mergers, are still very expected. 
For the tensorial GWs from binary mergers, it is already known that the proportions of two polarizations depend on the orbital inclination $\iota$ (angle between the sight direction and the spin axis of the binary)\cite{PhysRevLett.116.241102}. E.g., for the ``face-on'' ($\cos\iota=\pm1$) and ``edge-on''  ($\cos\iota=0$)  directions the GWs are circularly and linearly polarized, respectively. 
Excitingly, recent work\cite{PhysRevD.98.022008} presents the inclination-angle dependence and relative amplitudes for GWs including nontensorial modes, i.e., for modes of $h_x, h_y, h_b$ and $h_l$, the inclination angle $\iota$ gives factors of $\sin2\iota$, $\sin\iota$, $\sin^2\iota$ and $\sin2\iota$, respectively. Based on above knowledge,  we can have a specific manner of how the polarizations of perturbed ULF-EMWs connect to the tensorial and nontensorial polarizations of GWs from binary mergers. For a simple estimation, e.g.,  we here focus on the influence of inclination angle $\iota$ and ignore impact by other angular parameters, and also only include the vector modes as nontensorial GWs. According to the geometrical factors given by Ref.\cite{PhysRevD.98.022008}:
\begin{eqnarray}
\label{polarization1}
\mathcal{G}_{+}\propto\frac{5}{2}(1+cos^2\iota), ~\mathcal{G}_{\times}\propto i5\cos\iota,  \nonumber\\
\mathcal{G}_{V_x}\propto \sqrt{\frac{525}{56}}\sin2\iota, ~ \mathcal{G}_{V_y}\propto \sqrt{\frac{15}{2}}\sin\iota, 
\end{eqnarray}
for the $h_+$,   $h_{\times}$, $h_x$ and $h_y$ GWs, the mixed GWs can be expressed as\cite{PhysRevD.98.022008} (set the relative amplitudes of $h_+$ and $h_{\times}$ as 1 and equal to each other):
\begin{eqnarray}
\label{polarization2}
h=(\mathcal{G}_{+}+\mathcal{G}_{\times}+A_{V_x}\mathcal{G}_{V_x}+A_{V_y}\mathcal{G}_{V_y})h_{GR}, 
\end{eqnarray}
\indent On the other hand, due to current study\cite{Li.Wen.arXiv1712.00766}, the tensorial and nontensorial polarizations of GWs will lead to corresponding different polarizations of the perturbed ULF-EMWs, given particular types of background magnetic fields (transverse or longitudinal, to interact with the GWs). It is found\cite{Li.Wen.arXiv1712.00766}:\
\begin{eqnarray}
\label{polarization3}
\tilde{E}_x^{(1)}&\propto&{-h_{\times}B_x^{(0)}+h_{+}B_y^{(0)}-h_{y}B_z^{(0)}},\nonumber\\
\tilde{E}_y^{(1)}&\propto&{-h_{+}B_x^{(0)}+h_{\times}B_y^{(0)}+h_{x}B_z^{(0)}},
\end{eqnarray}
i.e., the electric component $\tilde{E}_x^{(1)}$ (in x-direction) of perturbed ULF-EMWs can be contributed by $h_{\times}$ interacting with $B_x^{(0)}$ (background magnetic fields in transverse direction), and by $h_{+}$ interacting with $B_y^{(0)}$ (also transverse background magnetic fields), and by nontensorial $h_{y}$ interacting with $B_z^{(0)}$ [longitudinal background magnetic fields (which only interact with nontensorial GWs) in z direction (also the propagating direction of the GWs)]\cite{Li.Wen.arXiv1712.00766}. With Eqs.(\ref{polarization1}) to (\ref{polarization3}) we can obtain the relationship how the polarizations of perturbed ULF-EMWs connect to the tensorial and nontensorial polarizations of GWs from binary mergers:
\begin{eqnarray}
\label{polarization4}
 \tilde{E}_x^{(1)}\propto &-&i5\cos\iota B_x^{(0)}\nonumber\\
  &+&\frac{5}{2}(1+\cos^2\iota) B_y^{(0)} - A_{V_y}\sqrt{\frac{15}{2}}\sin\iota B_z^{(0)},\nonumber\\
 \tilde{E}_y^{(1)}\propto &-&\frac{5}{2}(1+\cos^2\iota) B_x^{(0)}  \nonumber\\
  &+&i5\cos\iota B_y^{(0)} + A_{V_x} \sqrt{\frac{525}{56}}\sin2\iota B_z^{(0)},
\end{eqnarray}
based on the above expressions we have a brief picture of some examples of  polarizations of perturbed ULF-EMWs   shown in Fig. \ref{polarizations}. These figures indicate that the polarization of the EM counterparts of the perturbed ULF-EMWs obviously depends on not only the amplitudes of nontensorial GWs ($A_{V_x}$, $A_{V_y}$) and the inclination $\iota$, but also on the levels of background magnetic fields and their directions.\\
\indent Therefore, if any specific polarization of the perturbed ULF-EMWs would be captured and recognized, we could reversely extrapolate the possible combination of proportions of all polarizations (including nontensorial ones) of the GWs from binary mergers. Here, only some simplified cases are presented, and further studies considering more parameters to influence the polarizations of perturbed ULF-EMWs will be carefully and  detailedly addressed in other works.

\section{Modification of perturbed ULF-EMWs due to amplification process of magnetic fields of binary}
\label{section.waveforms}
\begin{figure*}[!htbp]
	\centerline{\includegraphics[width=7 in]{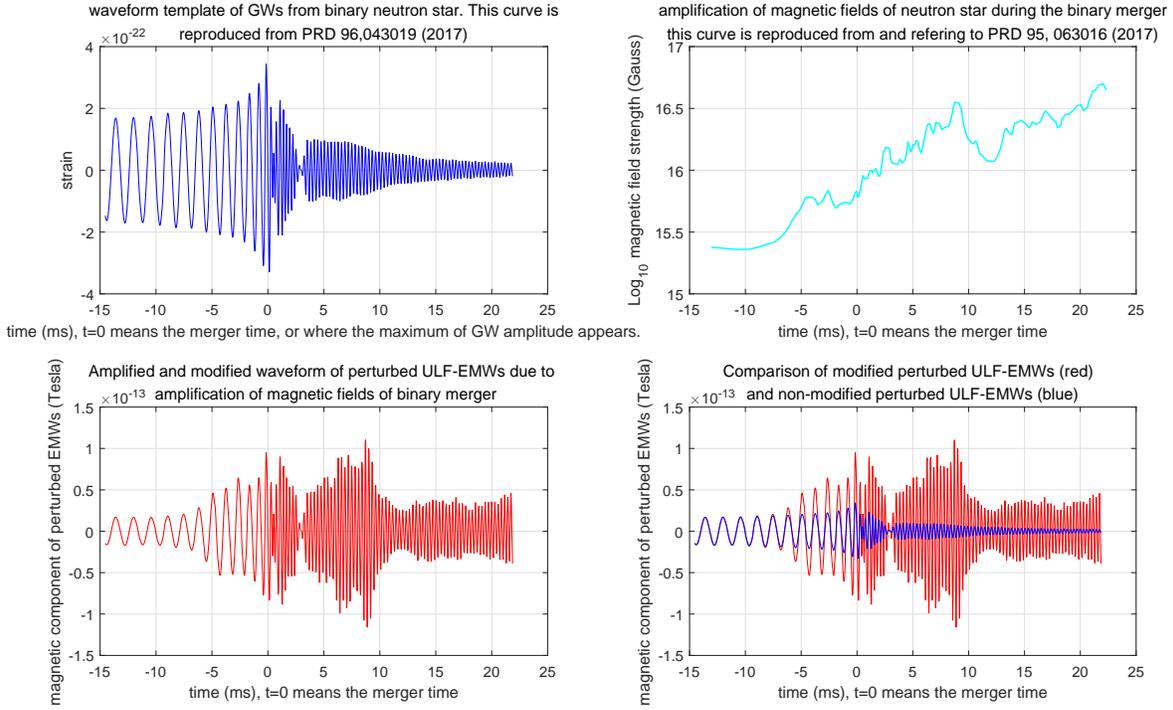}}
	\begin{spacing}{1.2}
		\caption{\footnotesize{\textbf{Simplified example of modification for the waveform of perturbed ULF-EMWs   caused by GWs from binary mergers with the amplification process of magnetic fields.}
				Subfigure (a) is a typical waveform template of binary neutron star merger, and this curve is produced and modified from some materials of Ref.\cite{PhysRevD.96.043019}. The subfigure (b) is an example of typical process of magnetic field amplification during binary merger (the curve is produced based on some materials of Ref.\cite{PhysRevD.95.063016}), and this extreme phenomenon could amplify the magnetic fields into very high level of $\sim10^{16}$Tesla or more. Here, the \mbox{time = 0} means the merger time where the maximum of GW amplitude appears. The subfigure (c) presents that the waveform of perturbed ULF-EMWs should be no longer linearly proportional to the waveform of the GWs of binary merger, because of the amplification process of magnetic fields of the binary (see more in Sect. \ref{section.waveforms}), and thus, in this example, the waveform of ULF-EMWs will be enlarged from about -10 ms to 22 ms. In other words, information of the amplification transfers into the characteristic shape of the waveform of perturbed ULF-EMWs. The subfigure (d) shows the comparison between the   perturbed ULF-EMWs with and without the influence of the amplification process. It should be emphasized that, this figure is not the exactly predicted waveform of perturbed ULF-EMWs,  but only to intuitively present the overall increasing tendency of the perturbed ULF-EMWs due to the amplification of magnetic field of the binary,  based on  very simplified considerations, and the precisely predicted waveforms of the perturbed ULF-EMWs need to be computed by massive numerical calculations and simulations, and such works will be carried out in separated articles as the next steps.}}
		\label{waveforms}
	\end{spacing}
\end{figure*}
\indent In this section we are not going to predict the precise waveforms of the perturbed ULF-EMWs, because such complicated tasks should be done by massive numerical computing in other consequent works. {\color{black}Thus, here, under simple considerations, we try to generally point out that the waveforms of perturbed ULF-EMWs are not linearly proportional to the GW waveforms, but instead, should be essentially modified due to the varying of the magnetic fields of the mergers.} Based on various mechanisms, the amplification process of magnetic fields of the binary mergers had been widely studied\cite{0264-9381-16-6-201,Price719,PhysRevD.95.063016,PhysRevD.90.041502,PhysRevD.92.124034,PhysRevD.97.124039,PhysRevD.92.084064,2041-8205-769-2-L29,0264-9381-33-16-164001,0004-637X-809-1-39} as one key feature to further understand the mergers. \\
\indent E.g., Rasio and Shapiro first pointed  that the Kelvin-Helmholtz (KH) instability would significantly
amplify the magnetic fields of the binary merger\cite{0264-9381-16-6-201}. \\
\indent Recent work of general relativistic magnetohydrodynamic simulations by Ciolfi et. al.\cite{PhysRevD.95.063016} indicate that the amplification process can lead to magnetic fields up to $10^{16}$G to $10^{17}$G by the effect of the Kelvin-Helmholtz instability. Subgrid modeling was also applied\cite{0004-637X-809-1-39} to find that the amplifications of up to 5 orders of magnitude are possible and the level of $10^{16}$G can be easily reached.\\
\indent Research of the turbulent amplification of magnetic fields in
local high-resolution simulations\cite{2041-8205-769-2-L29}, presented the magnetic fields $\sim 10^{16}$G throughout the merger duration of the neutron star binary. Another study\cite{PhysRevD.92.124034} on the magnetic-field amplification due to the Kelvin-Helmholtz instability also found  that there is an at least $10^3$ factor for the magnetic fields of binary neutron star mergers and it can easily reach  $10^{15}G$ or higher.\\
\indent Price and Rosswog\cite{Price719} argued that the magnetic fields of neutron star in binary mergers could be amplified by several orders of magnitude, and it is highly probably much stronger than $2\times10^{15}$G for realized cases in nature, and therefore the amplification may lead to the strongest magnetic fields in the Universe.\\
\indent In brief, many previous studies generally indicated the greatly amplified magnetic fields in the amplification procces during the binary merger.
Obtained results of Eqs. (\ref{B_dipole_tensor}) to (\ref{perturbedBfield}) indicate that the waveform of perturbed ULF-EMWs should be linearly proportional and similar to the waveforms of GWs of binary mergers, but such magnetic field amplification will influence the interaction of the GWs with the magnetic fields, and thus result in that the waveforms of the perturbed ULF-EMWs also have an amplification and modification in corresponding time duration. Here, we take the waveform of a typical waveform template\cite{PhysRevD.96.043019} of binary neutron star merger as an example, and the Fig. \ref{waveforms} shows that the waveform of the perturbed ULF-EMWs should be changed according to the amplification of magnetic fields of the mergers. Besides, many other works also predict various curves of such magnetic field amplifications by massive numerical calculations, and if apply those curves, the corresponding waveforms of perturbed ULF-EMWs would have diverse shapes and modifications depending on different models. \\
\indent Waveforms of the perturbed ULF-EMWs will be very characteristic features helpful for \mbox{extracting} the signals of perturbed ULF-EMWs from background noise, by similar signal processing methods applied for LIGO\&Virgo data using matched filtering based on waveform templates. {\color{black}For this purpose, the waveforms of perturbed ULF-EMWs need to be computed by numerical calculations (e.g., using  methods similar to those in some previous studies\cite{PhysRevD.94.064012,PhysRevD.96.043019,PhysRevD.95.063016,PhysRevD.91.044033,PhysRevD.92.044034}) as works carried out in future.}\\

\section{Summary and discussion}
\label{summary}
\textbf{Summary:}\\

\indent (1) Perturbed ultra low frequency EMWs ($\sim 10^2$ to $\sim 10^3$ Hz band, caused by GWs of binary mergers interacting with ultra-strong magnetic fields of neutron stars in the same binary) as a new type of signals and special EM counterparts  of GWs, have been addressed and estimated for various cases: tensorial-dipole [tensorial GWs interacting with dipole magnetic fields, Eq. (\ref{B_dipole_tensor})], nontensorial-dipole [Eq. (\ref{B_dipole_nontensor})], tensorial-quadrupole [Eq. (\ref{B_quad_tensor})] and nontensorial-quadrupole [Eq. (\ref{B_quad_nontensor})] cases. The strengths of magnetic components of these perturbed ULF-EMWs (not including influence of the ISM) can be around levels  {\color{black} around  $\sim10^{-12}$Tesla to $\sim10^{-17}$Tesla} at the Earth [Table \ref{perturbedBfarfield} and Fig.  \ref{Bcurves}] . Such levels would be approaching the sensitivity windows of ULF detectors and magnetometers	. \\
\indent (2) Specific relationship of how the polarizations of perturbed ULF-EMWs connect to the tensorial and possible nontensorial polarizations  of GWs from binary mergers, are addressed [Eq. \ref{polarization4} and Fig. \ref{polarizations}]. If such particular polarizations would be captured and recognized, we could reversely extrapolate the possible proportions of all polarizations (including nontensorial ones) of the GWs from binary mergers.\\
\indent (3) Due to the amplification process of magnetic fields of binary mergers, which can be greatly amplified by more than 3 order of magnitude and easily reaching  $10^{16}$G  or higher, the characteristic waveforms of the perturbed ULF-EMWs will be modified and have very unique shapes.
Inversely, the  ULF-EMWs may also provide useful information to investigate the evolution (including the amplification process) of the magnetic fields during the binary mergers.\\
\indent (4) The perturbed ULF-EMWs offer us a new type of EM counterparts of GWs from binary mergers. Such signals and EM counterparts have very characteristic waveforms and particular polarizations, and sit in totally different frequency band rather than usual EM counterpart of GRBs. \\
\indent Besides, importantly, for EM counterparts\cite{2041-8205-826-1-L6,AA2012,AA2016,Greiner,Smartt,0067-0049-225-1-8,2041-8205-826-2-L29,liuyuxiao,Nissanke,Kocsis,PhysRevD.81.084007,Lazzati,Lamb,0004-637X-835-1-103} such as GRBs, it is usually assumed that the EMWs and GWs are generated at the same time, but actually, there is still some unknown uncertainty (regarding the relative timing of emissions between GWs and EMWs)  would impact the analysis based on the difference of  arrival times between the EM and GW signals; differently, for the case of EM counterparts of the perturbed ULF-EMWs, such uncertainty would be reduced or avoided, because under the frame of electrodynamics in curved spacetime, they just clearly have the same start time to the GWs from the source of binary. Therefore, if there is some difference of arrival times between the GWs and ULF-EMWs, it would provide more accurate information underlying researches for some very important issues such as extra-dimensions of space, inflation, large-scale structure of Universe, measurement of cosmological parameters (e.g. local Hubble constant), speed and mass of photons and gravitons, Lorentz violations in gravity, and some other crucial properties of gravity and Universe.\\
%

\textbf{Discussion:}\\

\indent (1) If observation of the perturbed ULF-EMWs is possible, it would suggest a potential new way to observe the GWs from binary mergers, and such way has different effects to probe both tensorial and nontensorial polarizations of GWs, which relevant to fundamental issues of modified gravity, extra-dimensions of space and so on. The LIGO, Virgo and KAGRA, etc will form a observatory network to seek more information of polarizations (including possible additional ones) of GWs. However, the perturbed ULF-EMWs would reflect special connection relationship of polarizations to that of the GWs, and thus might provide different new information relevant to the possible nontensorial modes.\\
\indent (2) The signals of the perturbed ULF-EMWs could be complementary to (and cross-checked with) signals from other GW detections by LIGO, Virgo, KAGRA and so on.
As multi-messenger, it would be able to provide new information to reduce the uncertainty of the source positions, 
and also may allow a more relaxed parameter space.\\ 
\indent (3) The perturbed ULF-EMWs would bring signals of GWs in broader frequency bands up to several kHz or even higher, covering the GWs from last-inspiral phase, merger phase and post-merger phase of the binaries, that the LIGO and Virgo do not completely cover currently.  \\ 
%
\indent (4)  The calculation in this article can be also similarly applied to the single spinning magnetars with asymmetric mass distribution, etc.  Such continuous GWs will have different characteristic waveforms rather than the transient GWs of binary mergers, and the polarizations of corresponding perturbed ULF-EMWs could also have distinguishable features, depending on the polarizations (including both tensorial and nontensorial) of these continuous GWs and parameters such as the inclination angle.\\
\indent (5) Besides, the proposal addressed here for the binary neutron star, can also be applied to the cases involving possible primordial black holes (PBHs), i.g.,  binary system of a neutron star and a PBH. 
If the PBHs in such binary systems have small masses (less than 1 solar mass, then can be distinguished from the regular astrophysical black holes), and thus the binary of NS-PBH could produce GWs in much higher frequencies, leading to perturbed ULF-EMWs (or in higher frequencies, by interaction between the GWs of their mergers and the magnetic fields of neutron stars in the binary) above the frequency of ISM, i.e., be able to propagate until the Earth. If such ULF-EMWs could be observed, it may provide exclusive evidence of the PBHs, and we will address detailed works of relevant topics in subsequent articles.\\
\indent (6) The issues of perturbed ULF-EMWs caused by GWs proposed here, might suggest us a multi-disciplinary topic of interesting studies involving many aspects of the GWs, astronomy, astrophysics, EM counterparts, weak EMW detection, magnetometer, numerical GR, gravity, cosmology, extra-dimensions, etc.\\
%
\begin{acknowledgments}
This work is supported in part by National Natural Science Foundation of China (Grant No.11605015, No.11375279, No.11873001, No.11847301), Fundamental Research Funds for the Central Universities (Grant No.2019CDXYWL0029, No.106112017CDJXY300003, No.2019CDJDWL0005), Science and Technology Research Program of Chongqing Municipal Education Commission (Grant No. KJQN201800105), Natural Science Foundation Project of Chongqing cstc2018jcyjAX0767.     We greatly thank very valuable discussions and helps by  {Prof. Zhou-Jian Cao},  {Prof. Fang-Yu Li},   Prof.  B. Giacomazzo, {Prof. M.V. Romalis},  {Prof. I. Savukov},   {Prof. Z. Gruji{\'{c}}},   {Prof. C. Granata} and {Dr. O. Alem}. 
\end{acknowledgments}

\bibliography{WenReferenceData20180716}

\providecommand{\noopsort}[1]{}\providecommand{\singleletter}[1]{#1}%
\begin{thebibliography}{102}%
\makeatletter
\providecommand \@ifxundefined [1]{%
 \@ifx{#1\undefined}
}%
\providecommand \@ifnum [1]{%
 \ifnum #1\expandafter \@firstoftwo
 \else \expandafter \@secondoftwo
 \fi
}%
\providecommand \@ifx [1]{%
 \ifx #1\expandafter \@firstoftwo
 \else \expandafter \@secondoftwo
 \fi
}%
\providecommand \natexlab [1]{#1}%
\providecommand \enquote  [1]{``#1''}%
\providecommand \bibnamefont  [1]{#1}%
\providecommand \bibfnamefont [1]{#1}%
\providecommand \citenamefont [1]{#1}%
\providecommand \href@noop [0]{\@secondoftwo}%
\providecommand \href [0]{\begingroup \@sanitize@url \@href}%
\providecommand \@href[1]{\@@startlink{#1}\@@href}%
\providecommand \@@href[1]{\endgroup#1\@@endlink}%
\providecommand \@sanitize@url [0]{\catcode `\\12\catcode `\$12\catcode
  `\&12\catcode `\#12\catcode `\^12\catcode `\_12\catcode `\%12\relax}%
\providecommand \@@startlink[1]{}%
\providecommand \@@endlink[0]{}%
\providecommand \url  [0]{\begingroup\@sanitize@url \@url }%
\providecommand \@url [1]{\endgroup\@href {#1}{\urlprefix }}%
\providecommand \urlprefix  [0]{URL }%
\providecommand \Eprint [0]{\href }%
\providecommand \doibase [0]{http://dx.doi.org/}%
\providecommand \selectlanguage [0]{\@gobble}%
\providecommand \bibinfo  [0]{\@secondoftwo}%
\providecommand \bibfield  [0]{\@secondoftwo}%
\providecommand \translation [1]{[#1]}%
\providecommand \BibitemOpen [0]{}%
\providecommand \bibitemStop [0]{}%
\providecommand \bibitemNoStop [0]{.\EOS\space}%
\providecommand \EOS [0]{\spacefactor3000\relax}%
\providecommand \BibitemShut  [1]{\csname bibitem#1\endcsname}%
\let\auto@bib@innerbib\@empty
\bibitem [{\citenamefont {Abbott}\ \emph
  {et~al.}(2016{\natexlab{a}})\citenamefont {Abbott} \emph
  {et~al.}}]{PhysRevLett.116.061102}%
  \BibitemOpen
  \bibfield  {author} {\bibinfo {author} {\bibfnamefont {B.~P.}\ \bibnamefont
  {Abbott}} \emph {et~al.} (\bibinfo {collaboration} {LIGO Scientific
  Collaboration and Virgo Collaboration}),\ }\href@noop {} {\bibfield
  {journal} {\bibinfo  {journal} {Phys. Rev. Lett.}\ }\textbf {\bibinfo
  {volume} {116}},\ \bibinfo {pages} {061102} (\bibinfo {year}
  {2016}{\natexlab{a}})}\BibitemShut {NoStop}%
\bibitem [{\citenamefont {Abbott}\ \emph
  {et~al.}(2016{\natexlab{b}})\citenamefont {Abbott} \emph
  {et~al.}}]{secondLIGOGW}%
  \BibitemOpen
  \bibfield  {author} {\bibinfo {author} {\bibfnamefont {B.~P.}\ \bibnamefont
  {Abbott}} \emph {et~al.} (\bibinfo {collaboration} {LIGO Scientific
  Collaboration and Virgo Collaboration}),\ }\href@noop {} {\bibfield
  {journal} {\bibinfo  {journal} {Phys. Rev. Lett.}\ }\textbf {\bibinfo
  {volume} {116}},\ \bibinfo {pages} {241103} (\bibinfo {year}
  {2016}{\natexlab{b}})}\BibitemShut {NoStop}%
\bibitem [{\citenamefont {Abbott}\ \emph
  {et~al.}(2017{\natexlab{a}})\citenamefont {Abbott} \emph
  {et~al.}}]{PhysRevLett.118.221101}%
  \BibitemOpen
  \bibfield  {author} {\bibinfo {author} {\bibfnamefont {B.~P.}\ \bibnamefont
  {Abbott}} \emph {et~al.} (\bibinfo {collaboration} {LIGO Scientific
  Collaboration and Virgo Collaboration}),\ }\href@noop {} {\bibfield
  {journal} {\bibinfo  {journal} {Phys. Rev. Lett.}\ }\textbf {\bibinfo
  {volume} {118}},\ \bibinfo {pages} {221101} (\bibinfo {year}
  {2017}{\natexlab{a}})}\BibitemShut {NoStop}%
\bibitem [{\citenamefont {Abbott}\ \emph
  {et~al.}(2017{\natexlab{b}})\citenamefont {Abbott} \emph
  {et~al.}}]{GW170608}%
  \BibitemOpen
  \bibfield  {author} {\bibinfo {author} {\bibfnamefont {B.~P.}\ \bibnamefont
  {Abbott}} \emph {et~al.},\ }\href@noop {} {\bibfield  {journal} {\bibinfo
  {journal} {The Astrophysical Journal Letters}\ }\textbf {\bibinfo {volume}
  {851}},\ \bibinfo {pages} {L35} (\bibinfo {year}
  {2017}{\natexlab{b}})}\BibitemShut {NoStop}%
\bibitem [{\citenamefont {Abbott}\ \emph
  {et~al.}(2017{\natexlab{c}})\citenamefont {Abbott} \emph
  {et~al.}}]{GW170814}%
  \BibitemOpen
  \bibfield  {author} {\bibinfo {author} {\bibfnamefont {B.~P.}\ \bibnamefont
  {Abbott}} \emph {et~al.} (\bibinfo {collaboration} {LIGO Scientific
  Collaboration and Virgo Collaboration}),\ }\href@noop {} {\bibfield
  {journal} {\bibinfo  {journal} {Phys. Rev. Lett.}\ }\textbf {\bibinfo
  {volume} {119}},\ \bibinfo {pages} {141101} (\bibinfo {year}
  {2017}{\natexlab{c}})}\BibitemShut {NoStop}%
\bibitem [{\citenamefont {Abbott}\ \emph
  {et~al.}(2017{\natexlab{d}})\citenamefont {Abbott} \emph {et~al.}}]{bns}%
  \BibitemOpen
  \bibfield  {author} {\bibinfo {author} {\bibfnamefont {B.~P.}\ \bibnamefont
  {Abbott}} \emph {et~al.} (\bibinfo {collaboration} {LIGO Scientific
  Collaboration and Virgo Collaboration}),\ }\href@noop {} {\bibfield
  {journal} {\bibinfo  {journal} {Phys. Rev. Lett.}\ }\textbf {\bibinfo
  {volume} {119}},\ \bibinfo {pages} {161101} (\bibinfo {year}
  {2017}{\natexlab{d}})}\BibitemShut {NoStop}%
\bibitem [{\citenamefont {Abbott}\ \emph {et~al.}()\citenamefont {Abbott} \emph
  {et~al.}}]{LIGOnew1}%
  \BibitemOpen
  \bibfield  {author} {\bibinfo {author} {\bibfnamefont {B.~P.}\ \bibnamefont
  {Abbott}} \emph {et~al.} (\bibinfo {collaboration} {LIGO Scientific
  Collaboration and Virgo Collaboration}),\ }\href@noop {} {\bibinfo  {journal}
  {arXiv:1811.12907 [astro-ph.HE]}\ }\BibitemShut {NoStop}%
\bibitem [{\citenamefont {Abbott}\ \emph
  {et~al.}(2016{\natexlab{c}})\citenamefont {Abbott} \emph
  {et~al.}}]{PRX041015}%
  \BibitemOpen
\bibfield  {journal} {  }\bibfield  {author} {\bibinfo {author} {\bibfnamefont
  {B.~P.}\ \bibnamefont {Abbott}} \emph {et~al.} (\bibinfo {collaboration}
  {LIGO Scientific Collaboration and Virgo Collaboration}),\ }\href {\doibase
  10.1103/PhysRevX.6.041015} {\bibfield  {journal} {\bibinfo  {journal} {Phys.
  Rev. X}\ }\textbf {\bibinfo {volume} {6}},\ \bibinfo {pages} {041015}
  (\bibinfo {year} {2016}{\natexlab{c}})}\BibitemShut {NoStop}%
\bibitem [{\citenamefont {Connaughton}\ \emph {et~al.}(2016)\citenamefont
  {Connaughton}, \citenamefont {Burns}, \citenamefont {Goldstein} \emph
  {et~al.}}]{2041-8205-826-1-L6}%
  \BibitemOpen
  \bibfield  {author} {\bibinfo {author} {\bibfnamefont {V.}~\bibnamefont
  {Connaughton}}, \bibinfo {author} {\bibfnamefont {E.}~\bibnamefont {Burns}},
  \bibinfo {author} {\bibfnamefont {A.}~\bibnamefont {Goldstein}},  \emph
  {et~al.},\ }\href@noop {} {\bibfield  {journal} {\bibinfo  {journal} {The
  Astrophysical Journal Letters}\ }\textbf {\bibinfo {volume} {826}},\ \bibinfo
  {pages} {L6} (\bibinfo {year} {2016})}\BibitemShut {NoStop}%
\bibitem [{\citenamefont {Abadie}\ \emph {et~al.}(2012)\citenamefont {Abadie}
  \emph {et~al.}}]{AA2012}%
  \BibitemOpen
  \bibfield  {author} {\bibinfo {author} {\bibfnamefont {J.}~\bibnamefont
  {Abadie}} \emph {et~al.},\ }\href@noop {} {\bibfield  {journal} {\bibinfo
  {journal} {Astronomy \& Astrophysics}\ }\textbf {\bibinfo {volume} {539}},\
  \bibinfo {pages} {A124} (\bibinfo {year} {2012})}\BibitemShut {NoStop}%
\bibitem [{\citenamefont {Bagoly}\ \emph {et~al.}(2016)\citenamefont {Bagoly}
  \emph {et~al.}}]{AA2016}%
  \BibitemOpen
  \bibfield  {author} {\bibinfo {author} {\bibfnamefont {Z.}~\bibnamefont
  {Bagoly}} \emph {et~al.},\ }\href@noop {} {\bibfield  {journal} {\bibinfo
  {journal} {Astronomy \& Astrophysics}\ }\textbf {\bibinfo {volume} {593}},\
  \bibinfo {pages} {L10} (\bibinfo {year} {2016})}\BibitemShut {NoStop}%
\bibitem [{\citenamefont {Greiner}\ \emph {et~al.}(2016)\citenamefont
  {Greiner}, \citenamefont {Burgess}, \citenamefont {Savchenko},\ and\
  \citenamefont {Yu}}]{Greiner}%
  \BibitemOpen
  \bibfield  {author} {\bibinfo {author} {\bibfnamefont {J.}~\bibnamefont
  {Greiner}}, \bibinfo {author} {\bibfnamefont {J.~M.}\ \bibnamefont
  {Burgess}}, \bibinfo {author} {\bibfnamefont {V.}~\bibnamefont {Savchenko}},
  \ and\ \bibinfo {author} {\bibfnamefont {H.-F.}\ \bibnamefont {Yu}},\
  }\href@noop {} {\bibfield  {journal} {\bibinfo  {journal} {The Astrophysical
  Journal Letters}\ }\textbf {\bibinfo {volume} {827}},\ \bibinfo {pages} {L38}
  (\bibinfo {year} {2016})}\BibitemShut {NoStop}%
\bibitem [{\citenamefont {Smartt}\ \emph {et~al.}(2017)\citenamefont {Smartt}
  \emph {et~al.}}]{Smartt}%
  \BibitemOpen
  \bibfield  {author} {\bibinfo {author} {\bibfnamefont {S.~J.}\ \bibnamefont
  {Smartt}} \emph {et~al.},\ }\href@noop {} {\bibfield  {journal} {\bibinfo
  {journal} {Nature}\ }\textbf {\bibinfo {volume} {551}},\ \bibinfo {pages}
  {75} (\bibinfo {year} {2017})}\BibitemShut {NoStop}%
\bibitem [{\citenamefont {Abbott}\ \emph
  {et~al.}(2016{\natexlab{d}})\citenamefont {Abbott} \emph
  {et~al.}}]{0067-0049-225-1-8}%
  \BibitemOpen
  \bibfield  {author} {\bibinfo {author} {\bibfnamefont {B.~P.}\ \bibnamefont
  {Abbott}} \emph {et~al.},\ }\href@noop {} {\bibfield  {journal} {\bibinfo
  {journal} {The Astrophysical Journal Supplement Series}\ }\textbf {\bibinfo
  {volume} {225}},\ \bibinfo {pages} {8} (\bibinfo {year}
  {2016}{\natexlab{d}})}\BibitemShut {NoStop}%
\bibitem [{\citenamefont {Cowperthwaite}\ \emph {et~al.}(2016)\citenamefont
  {Cowperthwaite}, \citenamefont {Berger}, \citenamefont {Soares-Santos} \emph
  {et~al.}}]{2041-8205-826-2-L29}%
  \BibitemOpen
  \bibfield  {author} {\bibinfo {author} {\bibfnamefont {P.~S.}\ \bibnamefont
  {Cowperthwaite}}, \bibinfo {author} {\bibfnamefont {E.}~\bibnamefont
  {Berger}}, \bibinfo {author} {\bibfnamefont {M.}~\bibnamefont
  {Soares-Santos}},  \emph {et~al.},\ }\href@noop {} {\bibfield  {journal}
  {\bibinfo  {journal} {The Astrophysical Journal Letters}\ }\textbf {\bibinfo
  {volume} {826}},\ \bibinfo {pages} {L29} (\bibinfo {year}
  {2016})}\BibitemShut {NoStop}%
\bibitem [{\citenamefont {Yu}\ \emph {et~al.}(2017)\citenamefont {Yu},
  \citenamefont {Gu}, \citenamefont {Huang}, \citenamefont {Wang},
  \citenamefont {Meng},\ and\ \citenamefont {Liu}}]{liuyuxiao}%
  \BibitemOpen
  \bibfield  {author} {\bibinfo {author} {\bibfnamefont {H.}~\bibnamefont
  {Yu}}, \bibinfo {author} {\bibfnamefont {B.-M.}\ \bibnamefont {Gu}}, \bibinfo
  {author} {\bibfnamefont {F.~P.}\ \bibnamefont {Huang}}, \bibinfo {author}
  {\bibfnamefont {Y.-Q.}\ \bibnamefont {Wang}}, \bibinfo {author}
  {\bibfnamefont {X.-H.}\ \bibnamefont {Meng}}, \ and\ \bibinfo {author}
  {\bibfnamefont {Y.-X.}\ \bibnamefont {Liu}},\ }\href@noop {} {\bibfield
  {journal} {\bibinfo  {journal} {Journal of Cosmology and Astroparticle
  Physics}\ }\textbf {\bibinfo {volume} {2017}},\ \bibinfo {pages} {039}
  (\bibinfo {year} {2017})}\BibitemShut {NoStop}%
\bibitem [{\citenamefont {Nissanke}\ \emph {et~al.}(2013)\citenamefont
  {Nissanke}, \citenamefont {Kasliwal},\ and\ \citenamefont
  {Georgieva}}]{Nissanke}%
  \BibitemOpen
  \bibfield  {author} {\bibinfo {author} {\bibfnamefont {S.}~\bibnamefont
  {Nissanke}}, \bibinfo {author} {\bibfnamefont {M.}~\bibnamefont {Kasliwal}},
  \ and\ \bibinfo {author} {\bibfnamefont {A.}~\bibnamefont {Georgieva}},\
  }\href@noop {} {\bibfield  {journal} {\bibinfo  {journal} {The Astrophysical
  Journal}\ }\textbf {\bibinfo {volume} {767}},\ \bibinfo {pages} {124}
  (\bibinfo {year} {2013})}\BibitemShut {NoStop}%
\bibitem [{\citenamefont {Kocsis}\ \emph {et~al.}(2008)\citenamefont {Kocsis},
  \citenamefont {Haiman},\ and\ \citenamefont {Menou}}]{Kocsis}%
  \BibitemOpen
  \bibfield  {author} {\bibinfo {author} {\bibfnamefont {B.}~\bibnamefont
  {Kocsis}}, \bibinfo {author} {\bibfnamefont {Z.}~\bibnamefont {Haiman}}, \
  and\ \bibinfo {author} {\bibfnamefont {K.}~\bibnamefont {Menou}},\
  }\href@noop {} {\bibfield  {journal} {\bibinfo  {journal} {The Astrophysical
  Journal}\ }\textbf {\bibinfo {volume} {684}},\ \bibinfo {pages} {870}
  (\bibinfo {year} {2008})}\BibitemShut {NoStop}%
\bibitem [{\citenamefont {Palenzuela}\ \emph {et~al.}(2010)\citenamefont
  {Palenzuela}, \citenamefont {Lehner},\ and\ \citenamefont
  {Yoshida}}]{PhysRevD.81.084007}%
  \BibitemOpen
  \bibfield  {author} {\bibinfo {author} {\bibfnamefont {C.}~\bibnamefont
  {Palenzuela}}, \bibinfo {author} {\bibfnamefont {L.}~\bibnamefont {Lehner}},
  \ and\ \bibinfo {author} {\bibfnamefont {S.}~\bibnamefont {Yoshida}},\ }\href
  {\doibase 10.1103/PhysRevD.81.084007} {\bibfield  {journal} {\bibinfo
  {journal} {Phys. Rev. D}\ }\textbf {\bibinfo {volume} {81}},\ \bibinfo
  {pages} {084007} (\bibinfo {year} {2010})}\BibitemShut {NoStop}%
\bibitem [{\citenamefont {Lazzati}\ \emph {et~al.}(2017)\citenamefont
  {Lazzati}, \citenamefont {Deich}, \citenamefont {Morsony},\ and\
  \citenamefont {Workman}}]{Lazzati}%
  \BibitemOpen
  \bibfield  {author} {\bibinfo {author} {\bibfnamefont {D.}~\bibnamefont
  {Lazzati}}, \bibinfo {author} {\bibfnamefont {A.}~\bibnamefont {Deich}},
  \bibinfo {author} {\bibfnamefont {B.~J.}\ \bibnamefont {Morsony}}, \ and\
  \bibinfo {author} {\bibfnamefont {J.~C.}\ \bibnamefont {Workman}},\
  }\href@noop {} {\bibfield  {journal} {\bibinfo  {journal} {Monthly Notices of
  the Royal Astronomical Society}\ }\textbf {\bibinfo {volume} {471}},\
  \bibinfo {pages} {1652} (\bibinfo {year} {2017})}\BibitemShut {NoStop}%
\bibitem [{\citenamefont {Lamb}\ and\ \citenamefont {Kobayashi}(2017)}]{Lamb}%
  \BibitemOpen
  \bibfield  {author} {\bibinfo {author} {\bibfnamefont {G.~P.}\ \bibnamefont
  {Lamb}}\ and\ \bibinfo {author} {\bibfnamefont {S.}~\bibnamefont
  {Kobayashi}},\ }\href@noop {} {\bibfield  {journal} {\bibinfo  {journal}
  {Monthly Notices of the Royal Astronomical Society}\ }\textbf {\bibinfo
  {volume} {472}},\ \bibinfo {pages} {4953} (\bibinfo {year}
  {2017})}\BibitemShut {NoStop}%
\bibitem [{\citenamefont {Takahashi}(2017)}]{0004-637X-835-1-103}%
  \BibitemOpen
  \bibfield  {author} {\bibinfo {author} {\bibfnamefont {R.}~\bibnamefont
  {Takahashi}},\ }\href@noop {} {\bibfield  {journal} {\bibinfo  {journal} {The
  Astrophysical Journal}\ }\textbf {\bibinfo {volume} {835}},\ \bibinfo {pages}
  {103} (\bibinfo {year} {2017})}\BibitemShut {NoStop}%
\bibitem [{\citenamefont {Braginsky}\ \emph {et~al.}(1973)\citenamefont
  {Braginsky}, \citenamefont {Grishchuk}, \citenamefont {Doroshkevich},
  \citenamefont {Zeldovich}, \citenamefont {Novikov},\ and\ \citenamefont
  {Sazhin}}]{Braginsky.Grishchuk.1973}%
  \BibitemOpen
  \bibfield  {author} {\bibinfo {author} {\bibfnamefont {V.~B.}\ \bibnamefont
  {Braginsky}}, \bibinfo {author} {\bibfnamefont {L.~P.}\ \bibnamefont
  {Grishchuk}}, \bibinfo {author} {\bibfnamefont {A.~G.}\ \bibnamefont
  {Doroshkevich}}, \bibinfo {author} {\bibfnamefont {Y.~B.}\ \bibnamefont
  {Zeldovich}}, \bibinfo {author} {\bibfnamefont {I.~D.}\ \bibnamefont
  {Novikov}}, \ and\ \bibinfo {author} {\bibfnamefont {M.~V.}\ \bibnamefont
  {Sazhin}},\ }\href@noop {} {\bibfield  {journal} {\bibinfo  {journal} {Zh.
  Eksp. Teor. Fi}\ }\textbf {\bibinfo {volume} {65}},\ \bibinfo {pages} {1729}
  (\bibinfo {year} {1973})}\BibitemShut {NoStop}%
\bibitem [{\citenamefont {Boccaletti}\ \emph {et~al.}(1970)\citenamefont
  {Boccaletti}, \citenamefont {{V. De Sabbata}}, \citenamefont {Fortint},\ and\
  \citenamefont {Gualdi}}]{Boccaletti_NuovoCim70_1970}%
  \BibitemOpen
  \bibfield  {author} {\bibinfo {author} {\bibfnamefont {D.}~\bibnamefont
  {Boccaletti}}, \bibinfo {author} {\bibnamefont {{V. De Sabbata}}}, \bibinfo
  {author} {\bibfnamefont {P.}~\bibnamefont {Fortint}}, \ and\ \bibinfo
  {author} {\bibfnamefont {C.}~\bibnamefont {Gualdi}},\ }\href {\doibase
  10.1007/BF02710177} {\bibfield  {journal} {\bibinfo  {journal} {Nuovo Cim.
  B}\ }\textbf {\bibinfo {volume} {70}},\ \bibinfo {pages} {129} (\bibinfo
  {year} {1970})}\BibitemShut {NoStop}%
\bibitem [{\citenamefont {DeLogi}\ and\ \citenamefont
  {Mickelson}(1977)}]{prd2915}%
  \BibitemOpen
  \bibfield  {author} {\bibinfo {author} {\bibfnamefont {W.~K.}\ \bibnamefont
  {DeLogi}}\ and\ \bibinfo {author} {\bibfnamefont {A.~R.}\ \bibnamefont
  {Mickelson}},\ }\href@noop {} {\bibfield  {journal} {\bibinfo  {journal}
  {Phys. Rev. D}\ }\textbf {\bibinfo {volume} {16}},\ \bibinfo {pages} {2915}
  (\bibinfo {year} {1977})}\BibitemShut {NoStop}%
\bibitem [{\citenamefont {Chen}(1995)}]{Chen1995PRL}%
  \BibitemOpen
  \bibfield  {author} {\bibinfo {author} {\bibfnamefont {P.}~\bibnamefont
  {Chen}},\ }\href {\doibase 10.1103/PhysRevLett.74.634} {\bibfield  {journal}
  {\bibinfo  {journal} {Phys. Rev. Lett.}\ }\textbf {\bibinfo {volume} {74}},\
  \bibinfo {pages} {634} (\bibinfo {year} {1995})}\BibitemShut {NoStop}%
\bibitem [{\citenamefont {Chen}(1994)}]{Chen1994}%
  \BibitemOpen
  \bibfield  {author} {\bibinfo {author} {\bibfnamefont {P.}~\bibnamefont
  {Chen}},\ }\href@noop {} {\bibfield  {journal} {\bibinfo  {journal} {Stanford
  Linear Accelerator Center Report(SLAC-PUB-6666)}\ ,\ \bibinfo {pages} {379}}
  (\bibinfo {year} {1994})}\BibitemShut {NoStop}%
\bibitem [{\citenamefont {Long}\ \emph {et~al.}(1994)\citenamefont {Long},
  \citenamefont {Soa},\ and\ \citenamefont {Tuan}}]{LONG1994382}%
  \BibitemOpen
  \bibfield  {author} {\bibinfo {author} {\bibfnamefont {H.~N.}\ \bibnamefont
  {Long}}, \bibinfo {author} {\bibfnamefont {D.~V.}\ \bibnamefont {Soa}}, \
  and\ \bibinfo {author} {\bibfnamefont {T.~A.}\ \bibnamefont {Tuan}},\
  }\href@noop {} {\bibfield  {journal} {\bibinfo  {journal} {Physics Letters
  A}\ }\textbf {\bibinfo {volume} {186}},\ \bibinfo {pages} {382 } (\bibinfo
  {year} {1994})}\BibitemShut {NoStop}%
\bibitem [{\citenamefont {Li}\ \emph {et~al.}(2003)\citenamefont {Li},
  \citenamefont {Tang},\ and\ \citenamefont {Shi}}]{FYLi_PRD67_2003}%
  \BibitemOpen
  \bibfield  {author} {\bibinfo {author} {\bibfnamefont {F.~Y.}\ \bibnamefont
  {Li}}, \bibinfo {author} {\bibfnamefont {M.~X.}\ \bibnamefont {Tang}}, \ and\
  \bibinfo {author} {\bibfnamefont {D.~P.}\ \bibnamefont {Shi}},\ }\href@noop
  {} {\bibfield  {journal} {\bibinfo  {journal} {Phys. Rev. D}\ }\textbf
  {\bibinfo {volume} {67}},\ \bibinfo {pages} {104008} (\bibinfo {year}
  {2003})}\BibitemShut {NoStop}%
\bibitem [{\citenamefont {Li}\ \emph {et~al.}(2008)\citenamefont {Li},
  \citenamefont {{R. M. L. Baker, Jr.}}, \citenamefont {Fang}, \citenamefont
  {Stepheson},\ and\ \citenamefont {Chen}}]{FYLi_EPJC_2008}%
  \BibitemOpen
  \bibfield  {author} {\bibinfo {author} {\bibfnamefont {F.~Y.}\ \bibnamefont
  {Li}}, \bibinfo {author} {\bibnamefont {{R. M. L. Baker, Jr.}}}, \bibinfo
  {author} {\bibfnamefont {Z.~Y.}\ \bibnamefont {Fang}}, \bibinfo {author}
  {\bibfnamefont {G.~V.}\ \bibnamefont {Stepheson}}, \ and\ \bibinfo {author}
  {\bibfnamefont {Z.~Y.}\ \bibnamefont {Chen}},\ }\href@noop {} {\bibfield
  {journal} {\bibinfo  {journal} {Eur. Phys. J. C}\ }\textbf {\bibinfo {volume}
  {56}},\ \bibinfo {pages} {407} (\bibinfo {year} {2008})}\BibitemShut
  {NoStop}%
\bibitem [{\citenamefont {Li}\ \emph {et~al.}(2009)\citenamefont {Li},
  \citenamefont {Yang}, \citenamefont {Fang}, \citenamefont {Baker},
  \citenamefont {Stephenson},\ and\ \citenamefont {Wen}}]{FYLi_PRD80_2009}%
  \BibitemOpen
  \bibfield  {author} {\bibinfo {author} {\bibfnamefont {F.~Y.}\ \bibnamefont
  {Li}}, \bibinfo {author} {\bibfnamefont {N.}~\bibnamefont {Yang}}, \bibinfo
  {author} {\bibfnamefont {Z.~Y.}\ \bibnamefont {Fang}}, \bibinfo {author}
  {\bibfnamefont {R.~M.~L.}\ \bibnamefont {Baker}}, \bibinfo {author}
  {\bibfnamefont {G.~V.}\ \bibnamefont {Stephenson}}, \ and\ \bibinfo {author}
  {\bibfnamefont {H.}~\bibnamefont {Wen}},\ }\href@noop {} {\bibfield
  {journal} {\bibinfo  {journal} {Phys. Rev. D}\ }\textbf {\bibinfo {volume}
  {80}},\ \bibinfo {pages} {064013} (\bibinfo {year} {2009})}\BibitemShut
  {NoStop}%
\bibitem [{\citenamefont {Li}\ \emph {et~al.}(2013)\citenamefont {Li},
  \citenamefont {Wen},\ and\ \citenamefont {Fang}}]{Li.Fang-Yu.120402}%
  \BibitemOpen
  \bibfield  {author} {\bibinfo {author} {\bibfnamefont {F.~Y.}\ \bibnamefont
  {Li}}, \bibinfo {author} {\bibfnamefont {H.}~\bibnamefont {Wen}}, \ and\
  \bibinfo {author} {\bibfnamefont {Z.~Y.}\ \bibnamefont {Fang}},\ }\href@noop
  {} {\bibfield  {journal} {\bibinfo  {journal} {Chinese Physics B}\ }\textbf
  {\bibinfo {volume} {22}},\ \bibinfo {eid} {120402} (\bibinfo {year}
  {2013})}\BibitemShut {NoStop}%
\bibitem [{\citenamefont {Li}\ \emph {et~al.}(2011)\citenamefont {Li},
  \citenamefont {Lin}, \citenamefont {Li},\ and\ \citenamefont
  {Zhong}}]{Li2011}%
  \BibitemOpen
  \bibfield  {author} {\bibinfo {author} {\bibfnamefont {J.}~\bibnamefont
  {Li}}, \bibinfo {author} {\bibfnamefont {K.}~\bibnamefont {Lin}}, \bibinfo
  {author} {\bibfnamefont {F.}~\bibnamefont {Li}}, \ and\ \bibinfo {author}
  {\bibfnamefont {Y.}~\bibnamefont {Zhong}},\ }\href {\doibase
  10.1007/s10714-011-1176-8} {\bibfield  {journal} {\bibinfo  {journal}
  {General Relativity and Gravitation}\ }\textbf {\bibinfo {volume} {43}},\
  \bibinfo {pages} {2209} (\bibinfo {year} {2011})}\BibitemShut {NoStop}%
\bibitem [{\citenamefont {Wen}\ \emph {et~al.}(2019)\citenamefont {Wen},
  \citenamefont {Li}, \citenamefont {Li},\ and\ \citenamefont
  {Fang}}]{WEN2019114796}%
  \BibitemOpen
  \bibfield  {author} {\bibinfo {author} {\bibfnamefont {H.}~\bibnamefont
  {Wen}}, \bibinfo {author} {\bibfnamefont {F.-Y.}\ \bibnamefont {Li}},
  \bibinfo {author} {\bibfnamefont {J.}~\bibnamefont {Li}}, \ and\ \bibinfo
  {author} {\bibfnamefont {Z.-Y.}\ \bibnamefont {Fang}},\ }\href@noop {}
  {\bibfield  {journal} {\bibinfo  {journal} {Nuclear Physics B}\ }\textbf
  {\bibinfo {volume} {949}},\ \bibinfo {pages} {114796} (\bibinfo {year}
  {2019})}\BibitemShut {NoStop}%
\bibitem [{\citenamefont {Wen}\ \emph {et~al.}(2014{\natexlab{a}})\citenamefont
  {Wen}, \citenamefont {Li},\ and\ \citenamefont {Fang}}]{PRD104025}%
  \BibitemOpen
  \bibfield  {author} {\bibinfo {author} {\bibfnamefont {H.}~\bibnamefont
  {Wen}}, \bibinfo {author} {\bibfnamefont {F.~Y.}\ \bibnamefont {Li}}, \ and\
  \bibinfo {author} {\bibfnamefont {Z.~Y.}\ \bibnamefont {Fang}},\ }\href
  {\doibase 10.1103/PhysRevD.89.104025} {\bibfield  {journal} {\bibinfo
  {journal} {Phys. Rev. D}\ }\textbf {\bibinfo {volume} {89}},\ \bibinfo
  {pages} {104025} (\bibinfo {year} {2014}{\natexlab{a}})}\BibitemShut
  {NoStop}%
\bibitem [{\citenamefont {Wen}\ \emph {et~al.}(2014{\natexlab{b}})\citenamefont
  {Wen}, \citenamefont {Li}, \citenamefont {Fang},\ and\ \citenamefont
  {Beckwith}}]{WenEPJC2014}%
  \BibitemOpen
  \bibfield  {author} {\bibinfo {author} {\bibfnamefont {H.}~\bibnamefont
  {Wen}}, \bibinfo {author} {\bibfnamefont {F.~Y.}\ \bibnamefont {Li}},
  \bibinfo {author} {\bibfnamefont {Z.~Y.}\ \bibnamefont {Fang}}, \ and\
  \bibinfo {author} {\bibfnamefont {A.}~\bibnamefont {Beckwith}},\ }\href@noop
  {} {\bibfield  {journal} {\bibinfo  {journal} {The European Physical Journal
  C}\ }\textbf {\bibinfo {volume} {74}},\ \bibinfo {eid} {2998} (\bibinfo
  {year} {2014}{\natexlab{b}})}\BibitemShut {NoStop}%
\bibitem [{\citenamefont {Li}\ \emph {et~al.}(2016)\citenamefont {Li},
  \citenamefont {Wen}, \citenamefont {Fang}, \citenamefont {Wei}, \citenamefont
  {Wang},\ and\ \citenamefont {Zhang}}]{LiNPB2016}%
  \BibitemOpen
  \bibfield  {author} {\bibinfo {author} {\bibfnamefont {F.~Y.}\ \bibnamefont
  {Li}}, \bibinfo {author} {\bibfnamefont {H.}~\bibnamefont {Wen}}, \bibinfo
  {author} {\bibfnamefont {Z.~Y.}\ \bibnamefont {Fang}}, \bibinfo {author}
  {\bibfnamefont {L.~F.}\ \bibnamefont {Wei}}, \bibinfo {author} {\bibfnamefont
  {Y.~W.}\ \bibnamefont {Wang}}, \ and\ \bibinfo {author} {\bibfnamefont
  {M.}~\bibnamefont {Zhang}},\ }\href@noop {} {\bibfield  {journal} {\bibinfo
  {journal} {Nuclear Physics B}\ }\textbf {\bibinfo {volume} {911}},\ \bibinfo
  {pages} {500 } (\bibinfo {year} {2016})}\BibitemShut {NoStop}%
\bibitem [{\citenamefont {Wen}\ \emph {et~al.}(2017)\citenamefont {Wen},
  \citenamefont {Li}, \citenamefont {Li}, \citenamefont {Fang},\ and\
  \citenamefont {Beckwith}}]{wenCPC2017}%
  \BibitemOpen
  \bibfield  {author} {\bibinfo {author} {\bibfnamefont {H.}~\bibnamefont
  {Wen}}, \bibinfo {author} {\bibfnamefont {F.-Y.}\ \bibnamefont {Li}},
  \bibinfo {author} {\bibfnamefont {J.}~\bibnamefont {Li}}, \bibinfo {author}
  {\bibfnamefont {Z.-Y.}\ \bibnamefont {Fang}}, \ and\ \bibinfo {author}
  {\bibfnamefont {A.}~\bibnamefont {Beckwith}},\ }\href
  {http://stacks.iop.org/1674-1137/41/i=12/a=125101} {\bibfield  {journal}
  {\bibinfo  {journal} {Chinese Physics C}\ }\textbf {\bibinfo {volume} {41}},\
  \bibinfo {pages} {125101} (\bibinfo {year} {2017})}\BibitemShut {NoStop}%
\bibitem [{\citenamefont {Zheng}\ \emph {et~al.}(2018)\citenamefont {Zheng},
  \citenamefont {Wei}, \citenamefont {Wen},\ and\ \citenamefont
  {Li}}]{PhysRevD.98.064028}%
  \BibitemOpen
  \bibfield  {author} {\bibinfo {author} {\bibfnamefont {H.}~\bibnamefont
  {Zheng}}, \bibinfo {author} {\bibfnamefont {L.~F.}\ \bibnamefont {Wei}},
  \bibinfo {author} {\bibfnamefont {H.}~\bibnamefont {Wen}}, \ and\ \bibinfo
  {author} {\bibfnamefont {F.~Y.}\ \bibnamefont {Li}},\ }\href {\doibase
  10.1103/PhysRevD.98.064028} {\bibfield  {journal} {\bibinfo  {journal} {Phys.
  Rev. D}\ }\textbf {\bibinfo {volume} {98}},\ \bibinfo {pages} {064028}
  (\bibinfo {year} {2018})}\BibitemShut {NoStop}%
\bibitem [{\citenamefont {Li}\ \emph {et~al.}(2018)\citenamefont {Li},
  \citenamefont {Wen}, \citenamefont {Fang}, \citenamefont {Li},\ and\
  \citenamefont {Zhang}}]{Li.Wen.arXiv1712.00766}%
  \BibitemOpen
  \bibfield  {author} {\bibinfo {author} {\bibfnamefont {F.-Y.}\ \bibnamefont
  {Li}}, \bibinfo {author} {\bibfnamefont {H.}~\bibnamefont {Wen}}, \bibinfo
  {author} {\bibfnamefont {Z.-Y.}\ \bibnamefont {Fang}}, \bibinfo {author}
  {\bibfnamefont {D.}~\bibnamefont {Li}}, \ and\ \bibinfo {author}
  {\bibfnamefont {T.-J.}\ \bibnamefont {Zhang}},\ }\href@noop {} {\bibfield
  {journal} {\bibinfo  {journal} {arXiv:1712.00766 [gr-qc]}\ } (\bibinfo {year}
  {2018})}\BibitemShut {NoStop}%
\bibitem [{\citenamefont {Zhang}(2016)}]{PhysRevD.94.024048}%
  \BibitemOpen
  \bibfield  {author} {\bibinfo {author} {\bibfnamefont {F.}~\bibnamefont
  {Zhang}},\ }\href {\doibase 10.1103/PhysRevD.94.024048} {\bibfield  {journal}
  {\bibinfo  {journal} {Phys. Rev. D}\ }\textbf {\bibinfo {volume} {94}},\
  \bibinfo {pages} {024048} (\bibinfo {year} {2016})}\BibitemShut {NoStop}%
\bibitem [{\citenamefont {R\"adler}\ \emph {et~al.}(2001)\citenamefont
  {R\"adler}, \citenamefont {Fuchs}, \citenamefont {Geppert}, \citenamefont
  {Rheinhardt},\ and\ \citenamefont {Zannias}}]{PhysRevD.64.083008}%
  \BibitemOpen
  \bibfield  {author} {\bibinfo {author} {\bibfnamefont {K.-H.}\ \bibnamefont
  {R\"adler}}, \bibinfo {author} {\bibfnamefont {H.}~\bibnamefont {Fuchs}},
  \bibinfo {author} {\bibfnamefont {U.}~\bibnamefont {Geppert}}, \bibinfo
  {author} {\bibfnamefont {M.}~\bibnamefont {Rheinhardt}}, \ and\ \bibinfo
  {author} {\bibfnamefont {T.}~\bibnamefont {Zannias}},\ }\href {\doibase
  10.1103/PhysRevD.64.083008} {\bibfield  {journal} {\bibinfo  {journal} {Phys.
  Rev. D}\ }\textbf {\bibinfo {volume} {64}},\ \bibinfo {pages} {083008}
  (\bibinfo {year} {2001})}\BibitemShut {NoStop}%
\bibitem [{rom()}]{romalis}%
  \BibitemOpen
  \href@noop {} {\bibinfo  {journal}
  {http://physics.princeton.edu/romalis/magnetometer/}\ }\BibitemShut {NoStop}%
\bibitem [{\citenamefont {Taylor}\ \emph {et~al.}(2008)\citenamefont {Taylor},
  \citenamefont {Cappellaro}, \citenamefont {Childress}, \citenamefont {Jiang},
  \citenamefont {Budker}, \citenamefont {Hemmer}, \citenamefont {Yacoby},
  \citenamefont {Walsworth},\ and\ \citenamefont {Lukin}}]{Taylor}%
  \BibitemOpen
\bibfield  {journal} {  }\bibfield  {author} {\bibinfo {author} {\bibfnamefont
  {J.~M.}\ \bibnamefont {Taylor}}, \bibinfo {author} {\bibfnamefont
  {P.}~\bibnamefont {Cappellaro}}, \bibinfo {author} {\bibfnamefont
  {L.}~\bibnamefont {Childress}}, \bibinfo {author} {\bibfnamefont
  {L.}~\bibnamefont {Jiang}}, \bibinfo {author} {\bibfnamefont
  {D.}~\bibnamefont {Budker}}, \bibinfo {author} {\bibfnamefont {P.~R.}\
  \bibnamefont {Hemmer}}, \bibinfo {author} {\bibfnamefont {A.}~\bibnamefont
  {Yacoby}}, \bibinfo {author} {\bibfnamefont {R.}~\bibnamefont {Walsworth}}, \
  and\ \bibinfo {author} {\bibfnamefont {M.~D.}\ \bibnamefont {Lukin}},\
  }\href@noop {} {\bibfield  {journal} {\bibinfo  {journal} {Nature Physics}\
  }\textbf {\bibinfo {volume} {4}},\ \bibinfo {pages} {810} (\bibinfo {year}
  {2008})}\BibitemShut {NoStop}%
\bibitem [{\citenamefont {Budker}\ and\ \citenamefont
  {Romalis}(2007)}]{Budker}%
  \BibitemOpen
  \bibfield  {author} {\bibinfo {author} {\bibfnamefont {D.}~\bibnamefont
  {Budker}}\ and\ \bibinfo {author} {\bibfnamefont {M.}~\bibnamefont
  {Romalis}},\ }\href@noop {} {\bibfield  {journal} {\bibinfo  {journal}
  {Nature Physics}\ }\textbf {\bibinfo {volume} {3}},\ \bibinfo {pages} {227}
  (\bibinfo {year} {2007})}\BibitemShut {NoStop}%
\bibitem [{\citenamefont {Kominis}\ \emph {et~al.}(2003)\citenamefont
  {Kominis}, \citenamefont {Kornack}, \citenamefont {Allred},\ and\
  \citenamefont {Romalis}}]{Kominis}%
  \BibitemOpen
  \bibfield  {author} {\bibinfo {author} {\bibfnamefont {I.~K.}\ \bibnamefont
  {Kominis}}, \bibinfo {author} {\bibfnamefont {T.~W.}\ \bibnamefont
  {Kornack}}, \bibinfo {author} {\bibfnamefont {J.~C.}\ \bibnamefont {Allred}},
  \ and\ \bibinfo {author} {\bibfnamefont {M.~V.}\ \bibnamefont {Romalis}},\
  }\href {\doibase 10.1038/nature01484} {\bibfield  {journal} {\bibinfo
  {journal} {Nature}\ }\textbf {\bibinfo {volume} {422}},\ \bibinfo {pages}
  {596} (\bibinfo {year} {2003})}\BibitemShut {NoStop}%
\bibitem [{\citenamefont {Allred}\ \emph {et~al.}(2002)\citenamefont {Allred},
  \citenamefont {Lyman}, \citenamefont {Kornack},\ and\ \citenamefont
  {Romalis}}]{PhysRevLett.89.130801}%
  \BibitemOpen
  \bibfield  {author} {\bibinfo {author} {\bibfnamefont {J.~C.}\ \bibnamefont
  {Allred}}, \bibinfo {author} {\bibfnamefont {R.~N.}\ \bibnamefont {Lyman}},
  \bibinfo {author} {\bibfnamefont {T.~W.}\ \bibnamefont {Kornack}}, \ and\
  \bibinfo {author} {\bibfnamefont {M.~V.}\ \bibnamefont {Romalis}},\
  }\href@noop {} {\bibfield  {journal} {\bibinfo  {journal} {Phys. Rev. Lett.}\
  }\textbf {\bibinfo {volume} {89}},\ \bibinfo {pages} {130801} (\bibinfo
  {year} {2002})}\BibitemShut {NoStop}%
\bibitem [{\citenamefont {Drung}\ \emph {et~al.}(2007)\citenamefont {Drung},
  \citenamefont {Abmann}, \citenamefont {Beyer}, \citenamefont {Kirste},
  \citenamefont {Peters}, \citenamefont {Ruede},\ and\ \citenamefont
  {Schurig}}]{Drung}%
  \BibitemOpen
  \bibfield  {author} {\bibinfo {author} {\bibfnamefont {D.}~\bibnamefont
  {Drung}}, \bibinfo {author} {\bibfnamefont {C.}~\bibnamefont {Abmann}},
  \bibinfo {author} {\bibfnamefont {J.}~\bibnamefont {Beyer}}, \bibinfo
  {author} {\bibfnamefont {A.}~\bibnamefont {Kirste}}, \bibinfo {author}
  {\bibfnamefont {M.}~\bibnamefont {Peters}}, \bibinfo {author} {\bibfnamefont
  {F.}~\bibnamefont {Ruede}}, \ and\ \bibinfo {author} {\bibfnamefont
  {T.}~\bibnamefont {Schurig}},\ }\href@noop {} {\bibfield  {journal} {\bibinfo
   {journal} {IEEE Transactions on Applied Superconductivity}\ }\textbf
  {\bibinfo {volume} {17}},\ \bibinfo {pages} {699} (\bibinfo {year}
  {2007})}\BibitemShut {NoStop}%
\bibitem [{\citenamefont {Dang}\ \emph {et~al.}(2010)\citenamefont {Dang},
  \citenamefont {Maloof},\ and\ \citenamefont {Romalis}}]{Dang.Romalis.2010}%
  \BibitemOpen
  \bibfield  {author} {\bibinfo {author} {\bibfnamefont {H.~B.}\ \bibnamefont
  {Dang}}, \bibinfo {author} {\bibfnamefont {A.~C.}\ \bibnamefont {Maloof}}, \
  and\ \bibinfo {author} {\bibfnamefont {M.~V.}\ \bibnamefont {Romalis}},\
  }\href@noop {} {\bibfield  {journal} {\bibinfo  {journal} {Applied Physics
  Letters}\ }\textbf {\bibinfo {volume} {97}},\ \bibinfo {pages} {151110}
  (\bibinfo {year} {2010})}\BibitemShut {NoStop}%
\bibitem [{\citenamefont {Lee}\ \emph {et~al.}(2006)\citenamefont {Lee},
  \citenamefont {Sauer}, \citenamefont {Seltzer}, \citenamefont {Alem},\ and\
  \citenamefont {Romalis}}]{Lee}%
  \BibitemOpen
  \bibfield  {author} {\bibinfo {author} {\bibfnamefont {S.-K.}\ \bibnamefont
  {Lee}}, \bibinfo {author} {\bibfnamefont {K.~L.}\ \bibnamefont {Sauer}},
  \bibinfo {author} {\bibfnamefont {S.~J.}\ \bibnamefont {Seltzer}}, \bibinfo
  {author} {\bibfnamefont {O.}~\bibnamefont {Alem}}, \ and\ \bibinfo {author}
  {\bibfnamefont {M.~V.}\ \bibnamefont {Romalis}},\ }\href@noop {} {\bibfield
  {journal} {\bibinfo  {journal} {Applied Physics Letters}\ }\textbf {\bibinfo
  {volume} {89}},\ \bibinfo {pages} {214106} (\bibinfo {year}
  {2006})}\BibitemShut {NoStop}%
\bibitem [{\citenamefont {Sheng}\ \emph {et~al.}(2013)\citenamefont {Sheng},
  \citenamefont {Li}, \citenamefont {Dural},\ and\ \citenamefont
  {Romalis}}]{Sheng}%
  \BibitemOpen
  \bibfield  {author} {\bibinfo {author} {\bibfnamefont {D.}~\bibnamefont
  {Sheng}}, \bibinfo {author} {\bibfnamefont {S.}~\bibnamefont {Li}}, \bibinfo
  {author} {\bibfnamefont {N.}~\bibnamefont {Dural}}, \ and\ \bibinfo {author}
  {\bibfnamefont {M.~V.}\ \bibnamefont {Romalis}},\ }\href@noop {} {\bibfield
  {journal} {\bibinfo  {journal} {Phys. Rev. Lett.}\ }\textbf {\bibinfo
  {volume} {110}},\ \bibinfo {pages} {160802} (\bibinfo {year}
  {2013})}\BibitemShut {NoStop}%
\bibitem [{\citenamefont {Groeger}\ \emph {et~al.}(2006)\citenamefont
  {Groeger}, \citenamefont {Bison}, \citenamefont {Schenker}, \citenamefont
  {Wynands},\ and\ \citenamefont {Weis}}]{Groeger2006}%
  \BibitemOpen
  \bibfield  {author} {\bibinfo {author} {\bibfnamefont {S.}~\bibnamefont
  {Groeger}}, \bibinfo {author} {\bibfnamefont {G.}~\bibnamefont {Bison}},
  \bibinfo {author} {\bibfnamefont {J.-L.}\ \bibnamefont {Schenker}}, \bibinfo
  {author} {\bibfnamefont {R.}~\bibnamefont {Wynands}}, \ and\ \bibinfo
  {author} {\bibfnamefont {A.}~\bibnamefont {Weis}},\ }\href@noop {} {\bibfield
   {journal} {\bibinfo  {journal} {The European Physical Journal D - Atomic,
  Molecular, Optical and Plasma Physics}\ }\textbf {\bibinfo {volume} {38}},\
  \bibinfo {pages} {239} (\bibinfo {year} {2006})}\BibitemShut {NoStop}%
\bibitem [{\citenamefont {Mhaskar}\ \emph {et~al.}(2012)\citenamefont
  {Mhaskar}, \citenamefont {Knappe},\ and\ \citenamefont
  {Kitching}}]{Kitching}%
  \BibitemOpen
  \bibfield  {author} {\bibinfo {author} {\bibfnamefont {R.}~\bibnamefont
  {Mhaskar}}, \bibinfo {author} {\bibfnamefont {S.}~\bibnamefont {Knappe}}, \
  and\ \bibinfo {author} {\bibfnamefont {J.}~\bibnamefont {Kitching}},\
  }\href@noop {} {\bibfield  {journal} {\bibinfo  {journal} {Applied Physics
  Letters}\ }\textbf {\bibinfo {volume} {101}},\ \bibinfo {pages} {241105}
  (\bibinfo {year} {2012})}\BibitemShut {NoStop}%
\bibitem [{\citenamefont {Ito}\ \emph {et~al.}(2011)\citenamefont {Ito},
  \citenamefont {Ohnishi}, \citenamefont {Kamada},\ and\ \citenamefont
  {Kobayashi}}]{Ito}%
  \BibitemOpen
  \bibfield  {author} {\bibinfo {author} {\bibfnamefont {Y.}~\bibnamefont
  {Ito}}, \bibinfo {author} {\bibfnamefont {H.}~\bibnamefont {Ohnishi}},
  \bibinfo {author} {\bibfnamefont {K.}~\bibnamefont {Kamada}}, \ and\ \bibinfo
  {author} {\bibfnamefont {T.}~\bibnamefont {Kobayashi}},\ }\href@noop {}
  {\bibfield  {journal} {\bibinfo  {journal} {IEEE Transactions on Magnetics}\
  }\textbf {\bibinfo {volume} {47}},\ \bibinfo {pages} {3550} (\bibinfo {year}
  {2011})}\BibitemShut {NoStop}%
\bibitem [{\citenamefont {Cooper}\ \emph {et~al.}(2016)\citenamefont {Cooper},
  \citenamefont {Prescott}, \citenamefont {Matz}, \citenamefont {Sauer},
  \citenamefont {Dural}, \citenamefont {Romalis}, \citenamefont {Foley},
  \citenamefont {Kornack}, \citenamefont {Monti},\ and\ \citenamefont
  {Okamitsu}}]{PhysRevApplied.6.064014}%
  \BibitemOpen
  \bibfield  {author} {\bibinfo {author} {\bibfnamefont {R.~J.}\ \bibnamefont
  {Cooper}}, \bibinfo {author} {\bibfnamefont {D.~W.}\ \bibnamefont
  {Prescott}}, \bibinfo {author} {\bibfnamefont {P.}~\bibnamefont {Matz}},
  \bibinfo {author} {\bibfnamefont {K.~L.}\ \bibnamefont {Sauer}}, \bibinfo
  {author} {\bibfnamefont {N.}~\bibnamefont {Dural}}, \bibinfo {author}
  {\bibfnamefont {M.~V.}\ \bibnamefont {Romalis}}, \bibinfo {author}
  {\bibfnamefont {E.~L.}\ \bibnamefont {Foley}}, \bibinfo {author}
  {\bibfnamefont {T.~W.}\ \bibnamefont {Kornack}}, \bibinfo {author}
  {\bibfnamefont {M.}~\bibnamefont {Monti}}, \ and\ \bibinfo {author}
  {\bibfnamefont {J.}~\bibnamefont {Okamitsu}},\ }\href {\doibase
  10.1103/PhysRevApplied.6.064014} {\bibfield  {journal} {\bibinfo  {journal}
  {Phys. Rev. Applied}\ }\textbf {\bibinfo {volume} {6}},\ \bibinfo {pages}
  {064014} (\bibinfo {year} {2016})}\BibitemShut {NoStop}%
\bibitem [{\citenamefont {Breschi}\ \emph {et~al.}(2014)\citenamefont
  {Breschi}, \citenamefont {Gruji{\'{c}}}, \citenamefont {Knowles},\ and\
  \citenamefont {Weis}}]{Breschi}%
  \BibitemOpen
  \bibfield  {author} {\bibinfo {author} {\bibfnamefont {E.}~\bibnamefont
  {Breschi}}, \bibinfo {author} {\bibfnamefont {Z.~D.}\ \bibnamefont
  {Gruji{\'{c}}}}, \bibinfo {author} {\bibfnamefont {P.}~\bibnamefont
  {Knowles}}, \ and\ \bibinfo {author} {\bibfnamefont {A.}~\bibnamefont
  {Weis}},\ }\href@noop {} {\bibfield  {journal} {\bibinfo  {journal} {Applied
  Physics Letters}\ }\textbf {\bibinfo {volume} {104}},\ \bibinfo {pages}
  {023501} (\bibinfo {year} {2014})}\BibitemShut {NoStop}%
\bibitem [{\citenamefont {Prance}\ \emph {et~al.}(2003)\citenamefont {Prance},
  \citenamefont {Clark},\ and\ \citenamefont {Prance}}]{Prance}%
  \BibitemOpen
  \bibfield  {author} {\bibinfo {author} {\bibfnamefont {R.~J.}\ \bibnamefont
  {Prance}}, \bibinfo {author} {\bibfnamefont {T.~D.}\ \bibnamefont {Clark}}, \
  and\ \bibinfo {author} {\bibfnamefont {H.}~\bibnamefont {Prance}},\
  }\href@noop {} {\bibfield  {journal} {\bibinfo  {journal} {Review of
  Scientific Instruments}\ }\textbf {\bibinfo {volume} {74}},\ \bibinfo {pages}
  {3735} (\bibinfo {year} {2003})}\BibitemShut {NoStop}%
\bibitem [{\citenamefont {Savukov}\ \emph {et~al.}(2014)\citenamefont
  {Savukov}, \citenamefont {Karaulanov},\ and\ \citenamefont
  {Boshier}}]{Savukov}%
  \BibitemOpen
  \bibfield  {author} {\bibinfo {author} {\bibfnamefont {I.}~\bibnamefont
  {Savukov}}, \bibinfo {author} {\bibfnamefont {T.}~\bibnamefont {Karaulanov}},
  \ and\ \bibinfo {author} {\bibfnamefont {M.~G.}\ \bibnamefont {Boshier}},\
  }\href@noop {} {\bibfield  {journal} {\bibinfo  {journal} {Applied Physics
  Letters}\ }\textbf {\bibinfo {volume} {104}},\ \bibinfo {pages} {023504}
  (\bibinfo {year} {2014})}\BibitemShut {NoStop}%
\bibitem [{\citenamefont {Granata}\ \emph {et~al.}(2007)\citenamefont
  {Granata}, \citenamefont {Vettoliere},\ and\ \citenamefont
  {Russo}}]{Granata}%
  \BibitemOpen
  \bibfield  {author} {\bibinfo {author} {\bibfnamefont {C.}~\bibnamefont
  {Granata}}, \bibinfo {author} {\bibfnamefont {A.}~\bibnamefont {Vettoliere}},
  \ and\ \bibinfo {author} {\bibfnamefont {M.}~\bibnamefont {Russo}},\
  }\href@noop {} {\bibfield  {journal} {\bibinfo  {journal} {Applied Physics
  Letters}\ }\textbf {\bibinfo {volume} {91}},\ \bibinfo {pages} {122509}
  (\bibinfo {year} {2007})}\BibitemShut {NoStop}%
\bibitem [{\citenamefont {Gruji{\'{c}}}\ \emph {et~al.}(2015)\citenamefont
  {Gruji{\'{c}}}, \citenamefont {Koss}, \citenamefont {Bison},\ and\
  \citenamefont {Weis}}]{Grujic2015}%
  \BibitemOpen
  \bibfield  {author} {\bibinfo {author} {\bibfnamefont {Z.~D.}\ \bibnamefont
  {Gruji{\'{c}}}}, \bibinfo {author} {\bibfnamefont {P.~A.}\ \bibnamefont
  {Koss}}, \bibinfo {author} {\bibfnamefont {G.}~\bibnamefont {Bison}}, \ and\
  \bibinfo {author} {\bibfnamefont {A.}~\bibnamefont {Weis}},\ }\href@noop {}
  {\bibfield  {journal} {\bibinfo  {journal} {The European Physical Journal D}\
  }\textbf {\bibinfo {volume} {69}},\ \bibinfo {pages} {135} (\bibinfo {year}
  {2015})}\BibitemShut {NoStop}%
\bibitem [{\citenamefont {Lucivero}\ \emph {et~al.}(2014)\citenamefont
  {Lucivero}, \citenamefont {Anielski}, \citenamefont {Gawlik},\ and\
  \citenamefont {Mitchell}}]{Lucivero}%
  \BibitemOpen
  \bibfield  {author} {\bibinfo {author} {\bibfnamefont {V.~G.}\ \bibnamefont
  {Lucivero}}, \bibinfo {author} {\bibfnamefont {P.}~\bibnamefont {Anielski}},
  \bibinfo {author} {\bibfnamefont {W.}~\bibnamefont {Gawlik}}, \ and\ \bibinfo
  {author} {\bibfnamefont {M.~W.}\ \bibnamefont {Mitchell}},\ }\href@noop {}
  {\bibfield  {journal} {\bibinfo  {journal} {Review of Scientific
  Instruments}\ }\textbf {\bibinfo {volume} {85}},\ \bibinfo {pages} {113108}
  (\bibinfo {year} {2014})}\BibitemShut {NoStop}%
\bibitem [{\citenamefont {Alem}\ \emph {et~al.}(2013)\citenamefont {Alem},
  \citenamefont {Sauer},\ and\ \citenamefont {Romalis}}]{PhysRevA.87.013413}%
  \BibitemOpen
  \bibfield  {author} {\bibinfo {author} {\bibfnamefont {O.}~\bibnamefont
  {Alem}}, \bibinfo {author} {\bibfnamefont {K.~L.}\ \bibnamefont {Sauer}}, \
  and\ \bibinfo {author} {\bibfnamefont {M.~V.}\ \bibnamefont {Romalis}},\
  }\href@noop {} {\bibfield  {journal} {\bibinfo  {journal} {Phys. Rev. A}\
  }\textbf {\bibinfo {volume} {87}},\ \bibinfo {pages} {013413} (\bibinfo
  {year} {2013})}\BibitemShut {NoStop}%
\bibitem [{\citenamefont {Kulak}\ \emph {et~al.}(2014)\citenamefont {Kulak},
  \citenamefont {Kubisz}, \citenamefont {Klucjasz}, \citenamefont {Michalec},
  \citenamefont {Mlynarczyk}, \citenamefont {Nieckarz}, \citenamefont
  {Ostrowski},\ and\ \citenamefont {Zieba}}]{Kulak}%
  \BibitemOpen
  \bibfield  {author} {\bibinfo {author} {\bibfnamefont {A.}~\bibnamefont
  {Kulak}}, \bibinfo {author} {\bibfnamefont {J.}~\bibnamefont {Kubisz}},
  \bibinfo {author} {\bibfnamefont {S.}~\bibnamefont {Klucjasz}}, \bibinfo
  {author} {\bibfnamefont {A.}~\bibnamefont {Michalec}}, \bibinfo {author}
  {\bibfnamefont {J.}~\bibnamefont {Mlynarczyk}}, \bibinfo {author}
  {\bibfnamefont {Z.}~\bibnamefont {Nieckarz}}, \bibinfo {author}
  {\bibfnamefont {M.}~\bibnamefont {Ostrowski}}, \ and\ \bibinfo {author}
  {\bibfnamefont {S.}~\bibnamefont {Zieba}},\ }\href {\doibase
  10.1002/2014RS005400} {\bibfield  {journal} {\bibinfo  {journal} {Radio
  Science}\ }\textbf {\bibinfo {volume} {49}},\ \bibinfo {pages} {361}
  (\bibinfo {year} {2014})}\BibitemShut {NoStop}%
\bibitem [{\citenamefont {Balynsky}\ \emph {et~al.}(2017)\citenamefont
  {Balynsky}, \citenamefont {Gutierrez}, \citenamefont {Chiang}, \citenamefont
  {Kozhevnikov}, \citenamefont {Dudko}, \citenamefont {Filimonov},
  \citenamefont {Balandin},\ and\ \citenamefont {Khitun}}]{Balynsky}%
  \BibitemOpen
  \bibfield  {author} {\bibinfo {author} {\bibfnamefont {M.}~\bibnamefont
  {Balynsky}}, \bibinfo {author} {\bibfnamefont {D.}~\bibnamefont {Gutierrez}},
  \bibinfo {author} {\bibfnamefont {H.}~\bibnamefont {Chiang}}, \bibinfo
  {author} {\bibfnamefont {A.}~\bibnamefont {Kozhevnikov}}, \bibinfo {author}
  {\bibfnamefont {G.}~\bibnamefont {Dudko}}, \bibinfo {author} {\bibfnamefont
  {Y.}~\bibnamefont {Filimonov}}, \bibinfo {author} {\bibfnamefont {A.~A.}\
  \bibnamefont {Balandin}}, \ and\ \bibinfo {author} {\bibfnamefont
  {A.}~\bibnamefont {Khitun}},\ }\href@noop {} {\bibfield  {journal} {\bibinfo
  {journal} {Scientific Reports}\ ,\ \bibinfo {pages} {11539}} (\bibinfo {year}
  {2017})}\BibitemShut {NoStop}%
\bibitem [{\citenamefont {Limes}\ \emph {et~al.}(2018)\citenamefont {Limes},
  \citenamefont {Sheng},\ and\ \citenamefont
  {Romalis}}]{PhysRevLett.120.033401}%
  \BibitemOpen
  \bibfield  {author} {\bibinfo {author} {\bibfnamefont {M.~E.}\ \bibnamefont
  {Limes}}, \bibinfo {author} {\bibfnamefont {D.}~\bibnamefont {Sheng}}, \ and\
  \bibinfo {author} {\bibfnamefont {M.~V.}\ \bibnamefont {Romalis}},\ }\href
  {\doibase 10.1103/PhysRevLett.120.033401} {\bibfield  {journal} {\bibinfo
  {journal} {Phys. Rev. Lett.}\ }\textbf {\bibinfo {volume} {120}},\ \bibinfo
  {pages} {033401} (\bibinfo {year} {2018})}\BibitemShut {NoStop}%
\bibitem [{\citenamefont {Rasio}\ and\ \citenamefont
  {Shapiro}(1999)}]{0264-9381-16-6-201}%
  \BibitemOpen
  \bibfield  {author} {\bibinfo {author} {\bibfnamefont {F.~A.}\ \bibnamefont
  {Rasio}}\ and\ \bibinfo {author} {\bibfnamefont {S.~L.}\ \bibnamefont
  {Shapiro}},\ }\href@noop {} {\bibfield  {journal} {\bibinfo  {journal}
  {Classical and Quantum Gravity}\ }\textbf {\bibinfo {volume} {16}},\ \bibinfo
  {pages} {R1} (\bibinfo {year} {1999})}\BibitemShut {NoStop}%
\bibitem [{\citenamefont {Price}\ and\ \citenamefont
  {Rosswog}(2006)}]{Price719}%
  \BibitemOpen
  \bibfield  {author} {\bibinfo {author} {\bibfnamefont {D.~J.}\ \bibnamefont
  {Price}}\ and\ \bibinfo {author} {\bibfnamefont {S.}~\bibnamefont
  {Rosswog}},\ }\href@noop {} {\bibfield  {journal} {\bibinfo  {journal}
  {Science}\ }\textbf {\bibinfo {volume} {312}},\ \bibinfo {pages} {719}
  (\bibinfo {year} {2006})}\BibitemShut {NoStop}%
\bibitem [{\citenamefont {Ciolfi}\ \emph {et~al.}(2017)\citenamefont {Ciolfi},
  \citenamefont {Kastaun}, \citenamefont {Giacomazzo}, \citenamefont
  {Endrizzi}, \citenamefont {Siegel},\ and\ \citenamefont
  {Perna}}]{PhysRevD.95.063016}%
  \BibitemOpen
  \bibfield  {author} {\bibinfo {author} {\bibfnamefont {R.}~\bibnamefont
  {Ciolfi}}, \bibinfo {author} {\bibfnamefont {W.}~\bibnamefont {Kastaun}},
  \bibinfo {author} {\bibfnamefont {B.}~\bibnamefont {Giacomazzo}}, \bibinfo
  {author} {\bibfnamefont {A.}~\bibnamefont {Endrizzi}}, \bibinfo {author}
  {\bibfnamefont {D.~M.}\ \bibnamefont {Siegel}}, \ and\ \bibinfo {author}
  {\bibfnamefont {R.}~\bibnamefont {Perna}},\ }\href@noop {} {\bibfield
  {journal} {\bibinfo  {journal} {Phys. Rev. D}\ }\textbf {\bibinfo {volume}
  {95}},\ \bibinfo {pages} {063016} (\bibinfo {year} {2017})}\BibitemShut
  {NoStop}%
\bibitem [{\citenamefont {Kiuchi}\ \emph {et~al.}(2014)\citenamefont {Kiuchi},
  \citenamefont {Kyutoku}, \citenamefont {Sekiguchi}, \citenamefont {Shibata},\
  and\ \citenamefont {Wada}}]{PhysRevD.90.041502}%
  \BibitemOpen
  \bibfield  {author} {\bibinfo {author} {\bibfnamefont {K.}~\bibnamefont
  {Kiuchi}}, \bibinfo {author} {\bibfnamefont {K.}~\bibnamefont {Kyutoku}},
  \bibinfo {author} {\bibfnamefont {Y.}~\bibnamefont {Sekiguchi}}, \bibinfo
  {author} {\bibfnamefont {M.}~\bibnamefont {Shibata}}, \ and\ \bibinfo
  {author} {\bibfnamefont {T.}~\bibnamefont {Wada}},\ }\href {\doibase
  10.1103/PhysRevD.90.041502} {\bibfield  {journal} {\bibinfo  {journal} {Phys.
  Rev. D}\ }\textbf {\bibinfo {volume} {90}},\ \bibinfo {pages} {041502}
  (\bibinfo {year} {2014})}\BibitemShut {NoStop}%
\bibitem [{\citenamefont {Kiuchi}\ \emph {et~al.}(2015)\citenamefont {Kiuchi},
  \citenamefont {Cerd\'a-Dur\'an}, \citenamefont {Kyutoku}, \citenamefont
  {Sekiguchi},\ and\ \citenamefont {Shibata}}]{PhysRevD.92.124034}%
  \BibitemOpen
  \bibfield  {author} {\bibinfo {author} {\bibfnamefont {K.}~\bibnamefont
  {Kiuchi}}, \bibinfo {author} {\bibfnamefont {P.}~\bibnamefont
  {Cerd\'a-Dur\'an}}, \bibinfo {author} {\bibfnamefont {K.}~\bibnamefont
  {Kyutoku}}, \bibinfo {author} {\bibfnamefont {Y.}~\bibnamefont {Sekiguchi}},
  \ and\ \bibinfo {author} {\bibfnamefont {M.}~\bibnamefont {Shibata}},\
  }\href@noop {} {\bibfield  {journal} {\bibinfo  {journal} {Phys. Rev. D}\
  }\textbf {\bibinfo {volume} {92}},\ \bibinfo {pages} {124034} (\bibinfo
  {year} {2015})}\BibitemShut {NoStop}%
\bibitem [{\citenamefont {Kiuchi}\ \emph {et~al.}(2018)\citenamefont {Kiuchi},
  \citenamefont {Kyutoku}, \citenamefont {Sekiguchi},\ and\ \citenamefont
  {Shibata}}]{PhysRevD.97.124039}%
  \BibitemOpen
  \bibfield  {author} {\bibinfo {author} {\bibfnamefont {K.}~\bibnamefont
  {Kiuchi}}, \bibinfo {author} {\bibfnamefont {K.}~\bibnamefont {Kyutoku}},
  \bibinfo {author} {\bibfnamefont {Y.}~\bibnamefont {Sekiguchi}}, \ and\
  \bibinfo {author} {\bibfnamefont {M.}~\bibnamefont {Shibata}},\ }\href
  {\doibase 10.1103/PhysRevD.97.124039} {\bibfield  {journal} {\bibinfo
  {journal} {Phys. Rev. D}\ }\textbf {\bibinfo {volume} {97}},\ \bibinfo
  {pages} {124039} (\bibinfo {year} {2018})}\BibitemShut {NoStop}%
\bibitem [{\citenamefont {Dionysopoulou}\ \emph {et~al.}(2015)\citenamefont
  {Dionysopoulou}, \citenamefont {Alic},\ and\ \citenamefont
  {Rezzolla}}]{PhysRevD.92.084064}%
  \BibitemOpen
  \bibfield  {author} {\bibinfo {author} {\bibfnamefont {K.}~\bibnamefont
  {Dionysopoulou}}, \bibinfo {author} {\bibfnamefont {D.}~\bibnamefont {Alic}},
  \ and\ \bibinfo {author} {\bibfnamefont {L.}~\bibnamefont {Rezzolla}},\
  }\href {\doibase 10.1103/PhysRevD.92.084064} {\bibfield  {journal} {\bibinfo
  {journal} {Phys. Rev. D}\ }\textbf {\bibinfo {volume} {92}},\ \bibinfo
  {pages} {084064} (\bibinfo {year} {2015})}\BibitemShut {NoStop}%
\bibitem [{\citenamefont {Zrake}\ and\ \citenamefont
  {MacFadyen}(2013)}]{2041-8205-769-2-L29}%
  \BibitemOpen
  \bibfield  {author} {\bibinfo {author} {\bibfnamefont {J.}~\bibnamefont
  {Zrake}}\ and\ \bibinfo {author} {\bibfnamefont {A.~I.}\ \bibnamefont
  {MacFadyen}},\ }\href@noop {} {\bibfield  {journal} {\bibinfo  {journal} {The
  Astrophysical Journal Letters}\ }\textbf {\bibinfo {volume} {769}},\ \bibinfo
  {pages} {L29} (\bibinfo {year} {2013})}\BibitemShut {NoStop}%
\bibitem [{\citenamefont {Endrizzi}\ \emph {et~al.}(2016)\citenamefont
  {Endrizzi}, \citenamefont {Ciolfi}, \citenamefont {Giacomazzo}, \citenamefont
  {Kastaun},\ and\ \citenamefont {Kawamura}}]{0264-9381-33-16-164001}%
  \BibitemOpen
  \bibfield  {author} {\bibinfo {author} {\bibfnamefont {A.}~\bibnamefont
  {Endrizzi}}, \bibinfo {author} {\bibfnamefont {R.}~\bibnamefont {Ciolfi}},
  \bibinfo {author} {\bibfnamefont {B.}~\bibnamefont {Giacomazzo}}, \bibinfo
  {author} {\bibfnamefont {W.}~\bibnamefont {Kastaun}}, \ and\ \bibinfo
  {author} {\bibfnamefont {T.}~\bibnamefont {Kawamura}},\ }\href@noop {}
  {\bibfield  {journal} {\bibinfo  {journal} {Classical and Quantum Gravity}\
  }\textbf {\bibinfo {volume} {33}},\ \bibinfo {pages} {164001} (\bibinfo
  {year} {2016})}\BibitemShut {NoStop}%
\bibitem [{\citenamefont {Giacomazzo}\ \emph {et~al.}(2015)\citenamefont
  {Giacomazzo}, \citenamefont {Zrake}, \citenamefont {Duffell}, \citenamefont
  {MacFadyen},\ and\ \citenamefont {Perna}}]{0004-637X-809-1-39}%
  \BibitemOpen
  \bibfield  {author} {\bibinfo {author} {\bibfnamefont {B.}~\bibnamefont
  {Giacomazzo}}, \bibinfo {author} {\bibfnamefont {J.}~\bibnamefont {Zrake}},
  \bibinfo {author} {\bibfnamefont {P.~C.}\ \bibnamefont {Duffell}}, \bibinfo
  {author} {\bibfnamefont {A.~I.}\ \bibnamefont {MacFadyen}}, \ and\ \bibinfo
  {author} {\bibfnamefont {R.}~\bibnamefont {Perna}},\ }\href@noop {}
  {\bibfield  {journal} {\bibinfo  {journal} {The Astrophysical Journal}\
  }\textbf {\bibinfo {volume} {809}},\ \bibinfo {pages} {39} (\bibinfo {year}
  {2015})}\BibitemShut {NoStop}%
\bibitem [{\citenamefont {Eardley}\ \emph
  {et~al.}(1973{\natexlab{a}})\citenamefont {Eardley}, \citenamefont {Lee},
  \citenamefont {Lightman}, \citenamefont {Wagoner},\ and\ \citenamefont
  {Will}}]{PhysRevLett.30.884}%
  \BibitemOpen
  \bibfield  {author} {\bibinfo {author} {\bibfnamefont {D.~M.}\ \bibnamefont
  {Eardley}}, \bibinfo {author} {\bibfnamefont {D.~L.}\ \bibnamefont {Lee}},
  \bibinfo {author} {\bibfnamefont {A.~P.}\ \bibnamefont {Lightman}}, \bibinfo
  {author} {\bibfnamefont {R.~V.}\ \bibnamefont {Wagoner}}, \ and\ \bibinfo
  {author} {\bibfnamefont {C.~M.}\ \bibnamefont {Will}},\ }\href@noop {}
  {\bibfield  {journal} {\bibinfo  {journal} {Phys. Rev. Lett.}\ }\textbf
  {\bibinfo {volume} {30}},\ \bibinfo {pages} {884} (\bibinfo {year}
  {1973}{\natexlab{a}})}\BibitemShut {NoStop}%
\bibitem [{\citenamefont {Eardley}\ \emph
  {et~al.}(1973{\natexlab{b}})\citenamefont {Eardley}, \citenamefont {Lee},\
  and\ \citenamefont {Lightman}}]{PhysRevD.8.3308}%
  \BibitemOpen
  \bibfield  {author} {\bibinfo {author} {\bibfnamefont {D.~M.}\ \bibnamefont
  {Eardley}}, \bibinfo {author} {\bibfnamefont {D.~L.}\ \bibnamefont {Lee}}, \
  and\ \bibinfo {author} {\bibfnamefont {A.~P.}\ \bibnamefont {Lightman}},\
  }\href@noop {} {\bibfield  {journal} {\bibinfo  {journal} {Phys. Rev. D}\
  }\textbf {\bibinfo {volume} {8}},\ \bibinfo {pages} {3308} (\bibinfo {year}
  {1973}{\natexlab{b}})}\BibitemShut {NoStop}%
\bibitem [{\citenamefont {Takeda}\ \emph {et~al.}(2018)\citenamefont {Takeda},
  \citenamefont {Nishizawa}, \citenamefont {Michimura}, \citenamefont {Nagano},
  \citenamefont {Komori}, \citenamefont {Ando},\ and\ \citenamefont
  {Hayama}}]{PhysRevD.98.022008}%
  \BibitemOpen
  \bibfield  {author} {\bibinfo {author} {\bibfnamefont {H.}~\bibnamefont
  {Takeda}}, \bibinfo {author} {\bibfnamefont {A.}~\bibnamefont {Nishizawa}},
  \bibinfo {author} {\bibfnamefont {Y.}~\bibnamefont {Michimura}}, \bibinfo
  {author} {\bibfnamefont {K.}~\bibnamefont {Nagano}}, \bibinfo {author}
  {\bibfnamefont {K.}~\bibnamefont {Komori}}, \bibinfo {author} {\bibfnamefont
  {M.}~\bibnamefont {Ando}}, \ and\ \bibinfo {author} {\bibfnamefont
  {K.}~\bibnamefont {Hayama}},\ }\href@noop {} {\bibfield  {journal} {\bibinfo
  {journal} {Phys. Rev. D}\ }\textbf {\bibinfo {volume} {98}},\ \bibinfo
  {pages} {022008} (\bibinfo {year} {2018})}\BibitemShut {NoStop}%
\bibitem [{\citenamefont {Abbott}\ \emph {et~al.}(2012)\citenamefont {Abbott}
  \emph {et~al.}}]{LIGOAA}%
  \BibitemOpen
  \bibfield  {author} {\bibinfo {author} {\bibfnamefont {B.~P.}\ \bibnamefont
  {Abbott}} \emph {et~al.} (\bibinfo {collaboration} {LIGO Scientific
  Collaboration and Virgo Collaboration}),\ }\href@noop {} {\bibfield
  {journal} {\bibinfo  {journal} {A\&A}\ }\textbf {\bibinfo {volume} {539}},\
  \bibinfo {pages} {A124} (\bibinfo {year} {2012})}\BibitemShut {NoStop}%
\bibitem [{\citenamefont {Li}\ and\ \citenamefont
  {Paczy{\'{n}}ski}(1998)}]{Li_1998}%
  \BibitemOpen
  \bibfield  {author} {\bibinfo {author} {\bibfnamefont {L.-X.}\ \bibnamefont
  {Li}}\ and\ \bibinfo {author} {\bibfnamefont {B.}~\bibnamefont
  {Paczy{\'{n}}ski}},\ }\href {\doibase 10.1086/311680} {\bibfield  {journal}
  {\bibinfo  {journal} {The Astrophysical Journal}\ }\textbf {\bibinfo {volume}
  {507}},\ \bibinfo {pages} {L59} (\bibinfo {year} {1998})}\BibitemShut
  {NoStop}%
\bibitem [{\citenamefont {Thompson}(2008)}]{APJ.688.1258}%
  \BibitemOpen
  \bibfield  {author} {\bibinfo {author} {\bibfnamefont {C.}~\bibnamefont
  {Thompson}},\ }\href@noop {} {\bibfield  {journal} {\bibinfo  {journal} {The
  Astrophysical Journal}\ }\textbf {\bibinfo {volume} {688}},\ \bibinfo {pages}
  {1258} (\bibinfo {year} {2008})}\BibitemShut {NoStop}%
\bibitem [{\citenamefont {Gurnett}\ \emph {et~al.}(2013)\citenamefont
  {Gurnett}, \citenamefont {Kurth}, \citenamefont {Burlaga},\ and\
  \citenamefont {Ness}}]{Gurnett1489}%
  \BibitemOpen
  \bibfield  {author} {\bibinfo {author} {\bibfnamefont {D.~A.}\ \bibnamefont
  {Gurnett}}, \bibinfo {author} {\bibfnamefont {W.~S.}\ \bibnamefont {Kurth}},
  \bibinfo {author} {\bibfnamefont {L.~F.}\ \bibnamefont {Burlaga}}, \ and\
  \bibinfo {author} {\bibfnamefont {N.~F.}\ \bibnamefont {Ness}},\ }\href
  {\doibase 10.1126/science.1241681} {\bibfield  {journal} {\bibinfo  {journal}
  {Science}\ }\textbf {\bibinfo {volume} {341}},\ \bibinfo {pages} {1489}
  (\bibinfo {year} {2013})}\BibitemShut {NoStop}%
\bibitem [{\citenamefont {Gurnett}\ and\ \citenamefont
  {Bhattacharjee}(2005)}]{Gurnett2005}%
  \BibitemOpen
  \bibfield  {author} {\bibinfo {author} {\bibfnamefont {D.~A.}\ \bibnamefont
  {Gurnett}}\ and\ \bibinfo {author} {\bibfnamefont {A.}~\bibnamefont
  {Bhattacharjee}},\ }\href@noop {} {\emph {\bibinfo {title} {Introduction to
  plasma physics : with space, laboratory and astrophysical applications}}}\
  (\bibinfo  {publisher} {Cambridge University Press},\ \bibinfo {year}
  {2005})\BibitemShut {NoStop}%
\bibitem [{\citenamefont {Draine}(2010)}]{draine2010physics}%
  \BibitemOpen
  \bibfield  {author} {\bibinfo {author} {\bibfnamefont {B.}~\bibnamefont
  {Draine}},\ }\href {https://books.google.com/books?id=FycJvKHyiwsC} {\emph
  {\bibinfo {title} {Physics of the Interstellar and Intergalactic Medium}}},\
  Princeton Series in Astrophysics\ (\bibinfo  {publisher} {Princeton
  University Press},\ \bibinfo {year} {2010})\BibitemShut {NoStop}%
\bibitem [{\citenamefont {Zhang}\ \emph {et~al.}(2019)\citenamefont {Zhang},
  \citenamefont {Cao},\ and\ \citenamefont {Gao}}]{Cao2019IJMPD}%
  \BibitemOpen
  \bibfield  {author} {\bibinfo {author} {\bibfnamefont {X.}~\bibnamefont
  {Zhang}}, \bibinfo {author} {\bibfnamefont {Z.}~\bibnamefont {Cao}}, \ and\
  \bibinfo {author} {\bibfnamefont {H.}~\bibnamefont {Gao}},\ }\href {\doibase
  10.1142/S0218271819500263} {\bibfield  {journal} {\bibinfo  {journal}
  {International Journal of Modern Physics D}\ }\textbf {\bibinfo {volume}
  {28}},\ \bibinfo {pages} {1950026} (\bibinfo {year} {2019})},\ \Eprint
  {http://arxiv.org/abs/https://doi.org/10.1142/S0218271819500263}
  {https://doi.org/10.1142/S0218271819500263} \BibitemShut {NoStop}%
\bibitem [{\citenamefont {Shibata}\ and\ \citenamefont
  {Ury{\={u}}}(2002)}]{Shibata10.1143}%
  \BibitemOpen
  \bibfield  {author} {\bibinfo {author} {\bibfnamefont {M.}~\bibnamefont
  {Shibata}}\ and\ \bibinfo {author} {\bibfnamefont {K.}~\bibnamefont
  {Ury{\={u}}}},\ }\href {\doibase 10.1143/PTP.107.265} {\bibfield  {journal}
  {\bibinfo  {journal} {Progress of Theoretical Physics}\ }\textbf {\bibinfo
  {volume} {107}},\ \bibinfo {pages} {265} (\bibinfo {year} {2002})},\ \Eprint
  {http://arxiv.org/abs/http://oup.prod.sis.lan/ptp/article-pdf/107/2/265/5313264/107-2-265.pdf}
  {http://oup.prod.sis.lan/ptp/article-pdf/107/2/265/5313264/107-2-265.pdf}
  \BibitemShut {NoStop}%
\bibitem [{\citenamefont {Kiuchi}\ \emph {et~al.}(2009)\citenamefont {Kiuchi},
  \citenamefont {Sekiguchi}, \citenamefont {Shibata},\ and\ \citenamefont
  {Taniguchi}}]{PhysRevD.80.064037}%
  \BibitemOpen
  \bibfield  {author} {\bibinfo {author} {\bibfnamefont {K.}~\bibnamefont
  {Kiuchi}}, \bibinfo {author} {\bibfnamefont {Y.}~\bibnamefont {Sekiguchi}},
  \bibinfo {author} {\bibfnamefont {M.}~\bibnamefont {Shibata}}, \ and\
  \bibinfo {author} {\bibfnamefont {K.}~\bibnamefont {Taniguchi}},\ }\href
  {\doibase 10.1103/PhysRevD.80.064037} {\bibfield  {journal} {\bibinfo
  {journal} {Phys. Rev. D}\ }\textbf {\bibinfo {volume} {80}},\ \bibinfo
  {pages} {064037} (\bibinfo {year} {2009})}\BibitemShut {NoStop}%
\bibitem [{\citenamefont {Sekiguchi}\ \emph {et~al.}(2011)\citenamefont
  {Sekiguchi}, \citenamefont {Kiuchi}, \citenamefont {Kyutoku},\ and\
  \citenamefont {Shibata}}]{PhysRevLett.107.051102}%
  \BibitemOpen
  \bibfield  {author} {\bibinfo {author} {\bibfnamefont {Y.}~\bibnamefont
  {Sekiguchi}}, \bibinfo {author} {\bibfnamefont {K.}~\bibnamefont {Kiuchi}},
  \bibinfo {author} {\bibfnamefont {K.}~\bibnamefont {Kyutoku}}, \ and\
  \bibinfo {author} {\bibfnamefont {M.}~\bibnamefont {Shibata}},\ }\href
  {\doibase 10.1103/PhysRevLett.107.051102} {\bibfield  {journal} {\bibinfo
  {journal} {Phys. Rev. Lett.}\ }\textbf {\bibinfo {volume} {107}},\ \bibinfo
  {pages} {051102} (\bibinfo {year} {2011})}\BibitemShut {NoStop}%
\bibitem [{\citenamefont {Shibata}\ and\ \citenamefont
  {Ury\ifmmode~\bar{u}\else \={u}\fi{}}(2000)}]{PhysRevD.61.064001}%
  \BibitemOpen
  \bibfield  {author} {\bibinfo {author} {\bibfnamefont {M.}~\bibnamefont
  {Shibata}}\ and\ \bibinfo {author} {\bibfnamefont {K.~b.~o.}\ \bibnamefont
  {Ury\ifmmode~\bar{u}\else \={u}\fi{}}},\ }\href {\doibase
  10.1103/PhysRevD.61.064001} {\bibfield  {journal} {\bibinfo  {journal} {Phys.
  Rev. D}\ }\textbf {\bibinfo {volume} {61}},\ \bibinfo {pages} {064001}
  (\bibinfo {year} {2000})}\BibitemShut {NoStop}%
\bibitem [{\citenamefont {Oechslin}\ \emph {et~al.}(2002)\citenamefont
  {Oechslin}, \citenamefont {Rosswog},\ and\ \citenamefont
  {Thielemann}}]{PhysRevD.65.103005}%
  \BibitemOpen
  \bibfield  {author} {\bibinfo {author} {\bibfnamefont {R.}~\bibnamefont
  {Oechslin}}, \bibinfo {author} {\bibfnamefont {S.}~\bibnamefont {Rosswog}}, \
  and\ \bibinfo {author} {\bibfnamefont {F.-K.}\ \bibnamefont {Thielemann}},\
  }\href {\doibase 10.1103/PhysRevD.65.103005} {\bibfield  {journal} {\bibinfo
  {journal} {Phys. Rev. D}\ }\textbf {\bibinfo {volume} {65}},\ \bibinfo
  {pages} {103005} (\bibinfo {year} {2002})}\BibitemShut {NoStop}%
\bibitem [{\citenamefont {Zhuge}\ \emph {et~al.}(1994)\citenamefont {Zhuge},
  \citenamefont {Centrella},\ and\ \citenamefont
  {McMillan}}]{PhysRevD.50.6247}%
  \BibitemOpen
  \bibfield  {author} {\bibinfo {author} {\bibfnamefont {X.}~\bibnamefont
  {Zhuge}}, \bibinfo {author} {\bibfnamefont {J.~M.}\ \bibnamefont
  {Centrella}}, \ and\ \bibinfo {author} {\bibfnamefont {S.~L.~W.}\
  \bibnamefont {McMillan}},\ }\href {\doibase 10.1103/PhysRevD.50.6247}
  {\bibfield  {journal} {\bibinfo  {journal} {Phys. Rev. D}\ }\textbf {\bibinfo
  {volume} {50}},\ \bibinfo {pages} {6247} (\bibinfo {year}
  {1994})}\BibitemShut {NoStop}%
\bibitem [{\citenamefont {Faber}\ and\ \citenamefont
  {Rasio}(2012)}]{Faber2012}%
  \BibitemOpen
  \bibfield  {author} {\bibinfo {author} {\bibfnamefont {J.~A.}\ \bibnamefont
  {Faber}}\ and\ \bibinfo {author} {\bibfnamefont {F.~A.}\ \bibnamefont
  {Rasio}},\ }\href {\doibase 10.12942/lrr-2012-8} {\bibfield  {journal}
  {\bibinfo  {journal} {Living Reviews in Relativity}\ }\textbf {\bibinfo
  {volume} {15}},\ \bibinfo {pages} {8} (\bibinfo {year} {2012})}\BibitemShut
  {NoStop}%
\bibitem [{\citenamefont {Shibata}\ and\ \citenamefont
  {Ury\ifmmode~\bar{u}\else \={u}\fi{}}(2001)}]{PhysRevD.64.104017}%
  \BibitemOpen
  \bibfield  {author} {\bibinfo {author} {\bibfnamefont {M.}~\bibnamefont
  {Shibata}}\ and\ \bibinfo {author} {\bibfnamefont {K.~b.~o.}\ \bibnamefont
  {Ury\ifmmode~\bar{u}\else \={u}\fi{}}},\ }\href {\doibase
  10.1103/PhysRevD.64.104017} {\bibfield  {journal} {\bibinfo  {journal} {Phys.
  Rev. D}\ }\textbf {\bibinfo {volume} {64}},\ \bibinfo {pages} {104017}
  (\bibinfo {year} {2001})}\BibitemShut {NoStop}%
\bibitem [{\citenamefont {Abadie}\ \emph {et~al.}(2010)\citenamefont {Abadie},
  \citenamefont {Abbott}, \citenamefont {Abbott} \emph {et~al.}}]{Abadie_2010}%
  \BibitemOpen
  \bibfield  {author} {\bibinfo {author} {\bibfnamefont {J.}~\bibnamefont
  {Abadie}}, \bibinfo {author} {\bibfnamefont {B.~P.}\ \bibnamefont {Abbott}},
  \bibinfo {author} {\bibfnamefont {R.}~\bibnamefont {Abbott}},  \emph
  {et~al.},\ }\href {\doibase 10.1088/0264-9381/27/17/173001} {\bibfield
  {journal} {\bibinfo  {journal} {Classical and Quantum Gravity}\ }\textbf
  {\bibinfo {volume} {27}},\ \bibinfo {pages} {173001} (\bibinfo {year}
  {2010})}\BibitemShut {NoStop}%
\bibitem [{\citenamefont {Dietrich}\ \emph {et~al.}(2015)\citenamefont
  {Dietrich}, \citenamefont {Moldenhauer}, \citenamefont {Johnson-McDaniel},
  \citenamefont {Bernuzzi}, \citenamefont {Markakis}, \citenamefont
  {Br\"ugmann},\ and\ \citenamefont {Tichy}}]{PhysRevD.92.124007}%
  \BibitemOpen
  \bibfield  {author} {\bibinfo {author} {\bibfnamefont {T.}~\bibnamefont
  {Dietrich}}, \bibinfo {author} {\bibfnamefont {N.}~\bibnamefont
  {Moldenhauer}}, \bibinfo {author} {\bibfnamefont {N.~K.}\ \bibnamefont
  {Johnson-McDaniel}}, \bibinfo {author} {\bibfnamefont {S.}~\bibnamefont
  {Bernuzzi}}, \bibinfo {author} {\bibfnamefont {C.~M.}\ \bibnamefont
  {Markakis}}, \bibinfo {author} {\bibfnamefont {B.}~\bibnamefont
  {Br\"ugmann}}, \ and\ \bibinfo {author} {\bibfnamefont {W.}~\bibnamefont
  {Tichy}},\ }\href {\doibase 10.1103/PhysRevD.92.124007} {\bibfield  {journal}
  {\bibinfo  {journal} {Phys. Rev. D}\ }\textbf {\bibinfo {volume} {92}},\
  \bibinfo {pages} {124007} (\bibinfo {year} {2015})}\BibitemShut {NoStop}%
\bibitem [{\citenamefont {Abbott}\ \emph {et~al.}(2018)\citenamefont {Abbott}
  \emph {et~al.}}]{PRL120.031104}%
  \BibitemOpen
  \bibfield  {author} {\bibinfo {author} {\bibfnamefont {B.~P.}\ \bibnamefont
  {Abbott}} \emph {et~al.} (\bibinfo {collaboration} {LIGO Scientific
  Collaboration and Virgo Collaboration}),\ }\href@noop {} {\bibfield
  {journal} {\bibinfo  {journal} {Phys. Rev. Lett.}\ }\textbf {\bibinfo
  {volume} {120}},\ \bibinfo {pages} {031104} (\bibinfo {year}
  {2018})}\BibitemShut {NoStop}%
\bibitem [{\citenamefont {Isi}\ and\ \citenamefont
  {Weinstein}(2017)}]{arXiv1710.03794}%
  \BibitemOpen
  \bibfield  {author} {\bibinfo {author} {\bibfnamefont {M.}~\bibnamefont
  {Isi}}\ and\ \bibinfo {author} {\bibfnamefont {A.~J.}\ \bibnamefont
  {Weinstein}},\ }\href@noop {} {\bibfield  {journal} {\bibinfo  {journal}
  {arXiv:1710.03794 [gr-qc]}\ } (\bibinfo {year} {2017})}\BibitemShut {NoStop}%
\bibitem [{\citenamefont {Abbott}\ \emph
  {et~al.}(2016{\natexlab{e}})\citenamefont {Abbott} \emph
  {et~al.}}]{PhysRevLett.116.241102}%
  \BibitemOpen
  \bibfield  {author} {\bibinfo {author} {\bibfnamefont {B.~P.}\ \bibnamefont
  {Abbott}} \emph {et~al.} (\bibinfo {collaboration} {LIGO Scientific
  Collaboration and Virgo Collaboration}),\ }\href@noop {} {\bibfield
  {journal} {\bibinfo  {journal} {Phys. Rev. Lett.}\ }\textbf {\bibinfo
  {volume} {116}},\ \bibinfo {pages} {241102} (\bibinfo {year}
  {2016}{\natexlab{e}})}\BibitemShut {NoStop}%
\bibitem [{\citenamefont {Kastaun}\ \emph {et~al.}(2017)\citenamefont
  {Kastaun}, \citenamefont {Ciolfi}, \citenamefont {Endrizzi},\ and\
  \citenamefont {Giacomazzo}}]{PhysRevD.96.043019}%
  \BibitemOpen
  \bibfield  {author} {\bibinfo {author} {\bibfnamefont {W.}~\bibnamefont
  {Kastaun}}, \bibinfo {author} {\bibfnamefont {R.}~\bibnamefont {Ciolfi}},
  \bibinfo {author} {\bibfnamefont {A.}~\bibnamefont {Endrizzi}}, \ and\
  \bibinfo {author} {\bibfnamefont {B.}~\bibnamefont {Giacomazzo}},\ }\href
  {\doibase 10.1103/PhysRevD.96.043019} {\bibfield  {journal} {\bibinfo
  {journal} {Phys. Rev. D}\ }\textbf {\bibinfo {volume} {96}},\ \bibinfo
  {pages} {043019} (\bibinfo {year} {2017})}\BibitemShut {NoStop}%
\bibitem [{\citenamefont {Kawamura}\ \emph {et~al.}(2016)\citenamefont
  {Kawamura}, \citenamefont {Giacomazzo}, \citenamefont {Kastaun},
  \citenamefont {Ciolfi}, \citenamefont {Endrizzi}, \citenamefont {Baiotti},\
  and\ \citenamefont {Perna}}]{PhysRevD.94.064012}%
  \BibitemOpen
  \bibfield  {author} {\bibinfo {author} {\bibfnamefont {T.}~\bibnamefont
  {Kawamura}}, \bibinfo {author} {\bibfnamefont {B.}~\bibnamefont
  {Giacomazzo}}, \bibinfo {author} {\bibfnamefont {W.}~\bibnamefont {Kastaun}},
  \bibinfo {author} {\bibfnamefont {R.}~\bibnamefont {Ciolfi}}, \bibinfo
  {author} {\bibfnamefont {A.}~\bibnamefont {Endrizzi}}, \bibinfo {author}
  {\bibfnamefont {L.}~\bibnamefont {Baiotti}}, \ and\ \bibinfo {author}
  {\bibfnamefont {R.}~\bibnamefont {Perna}},\ }\href {\doibase
  10.1103/PhysRevD.94.064012} {\bibfield  {journal} {\bibinfo  {journal} {Phys.
  Rev. D}\ }\textbf {\bibinfo {volume} {94}},\ \bibinfo {pages} {064012}
  (\bibinfo {year} {2016})}\BibitemShut {NoStop}%
\bibitem [{\citenamefont {Cao}(2015)}]{PhysRevD.91.044033}%
  \BibitemOpen
  \bibfield  {author} {\bibinfo {author} {\bibfnamefont {Z.}~\bibnamefont
  {Cao}},\ }\href {\doibase 10.1103/PhysRevD.91.044033} {\bibfield  {journal}
  {\bibinfo  {journal} {Phys. Rev. D}\ }\textbf {\bibinfo {volume} {91}},\
  \bibinfo {pages} {044033} (\bibinfo {year} {2015})}\BibitemShut {NoStop}%
\bibitem [{\citenamefont {Sun}\ \emph {et~al.}(2015)\citenamefont {Sun},
  \citenamefont {Cao}, \citenamefont {Wang},\ and\ \citenamefont
  {Yeh}}]{PhysRevD.92.044034}%
  \BibitemOpen
  \bibfield  {author} {\bibinfo {author} {\bibfnamefont {B.}~\bibnamefont
  {Sun}}, \bibinfo {author} {\bibfnamefont {Z.}~\bibnamefont {Cao}}, \bibinfo
  {author} {\bibfnamefont {Y.}~\bibnamefont {Wang}}, \ and\ \bibinfo {author}
  {\bibfnamefont {H.-C.}\ \bibnamefont {Yeh}},\ }\href {\doibase
  10.1103/PhysRevD.92.044034} {\bibfield  {journal} {\bibinfo  {journal} {Phys.
  Rev. D}\ }\textbf {\bibinfo {volume} {92}},\ \bibinfo {pages} {044034}
  (\bibinfo {year} {2015})}\BibitemShut {NoStop}%
\end{thebibliography}%

\end{document}